\documentclass[twocolumn,showpacs,showkeys,amssymb,nobibnotes,superscriptaddress,aps,prd,bibnotes]{revtex4-2}

\usepackage{graphicx}
\usepackage{dcolumn}
\usepackage{bm}
\usepackage[T1]{fontenc}
\usepackage{mathptmx}
\usepackage{mathrsfs}
\usepackage{physics}
\usepackage{simplewick}
\usepackage{mathrsfs}
\usepackage{amsthm,amsmath,amssymb}

\usepackage[bottom]{footmisc}

\usepackage{amssymb}
\usepackage{amsmath}
\usepackage{amsfonts}
\usepackage{bm}
\usepackage{graphicx}
\usepackage{subfigure}
\usepackage[dvipsnames]{xcolor}
\usepackage{calc}
\usepackage{epsfig}
\usepackage{color}
\usepackage[colorlinks,citecolor=blue]{hyperref}
\begin{document}
\bibliographystyle{apsrev4-1}
\newcommand{\be}{\begin{equation}}
\newcommand{\ee}{\end{equation}}
\newcommand{\bs}{\begin{split}}
\newcommand{\es}{\end{split}}
\newcommand{\R}[1]{\textcolor{red}{#1}}
\newcommand{\B}[1]{\textcolor{blue}{#1}}

\title{Boosting the sensitivity of high frequency gravitational wave detectors by PT-symmetry}
\author{Chuming Wang}
\affiliation{Center for Gravitational Experiment, Hubei Key Laboratory of Gravitation and Quantum Physics, School of Physics, Huazhong University of Science and Technology, Wuhan, 430074, P. R. China}
\author{Chunnong Zhao}
\affiliation{ARC Centre of Excellence for Gravitational Wave Discovery, The University of Western Australia,
Crawley, Western Australia 6009, Australia}
\author{Xiang Li}
\affiliation{Burke Institute for Theoretical Physics, California Institute of Technology, Pasadena, California 91125, USA}
\author{Enping Zhou}
\affiliation{Department of Astronomy, School of Physics, Huazhong University of Science and Technology, Wuhan, 430074, P. R. China}
\author{Haixing Miao}
\affiliation{State Key Laboratory of Low Dimensional Quantum Physics, Department of Physics, Tsinghua University, Beijing, China}
\author{Yanbei Chen}
\affiliation{Burke Institute for Theoretical Physics, California Institute of Technology, Pasadena, California 91125, USA}
\author{Yiqiu Ma}
\email{myqphy@hust.edu.cn}
\affiliation{Center for Gravitational Experiment, Hubei Key Laboratory of Gravitation and Quantum Physics, School of Physics, Huazhong University of Science and Technology, Wuhan, 430074, P. R. China}
\affiliation{Department of Astronomy, School of Physics, Huazhong University of Science and Technology, Wuhan, 430074, P. R. China}

\begin{abstract}
The kilo-Hertz gravitational waves radiated by the neutron star merger remnants carry rich information about the physics of high-density nuclear matter states, and many important astrophysical phenomena such as gamma-ray bursts and black hole formation. Current laser interferometer gravitational wave detectors, such as LIGO, VIRGO, and KAGRA have limited signal response at the kilo-Hertz band, thereby unable to capture these important physical phenomena. This work proposes an alternative protocol for boosting the sensitivity of the gravitational wave detectors at high frequency by implementing an optomechanical quantum amplifier. With the auxiliary quantum amplifier, this design has the feature of Parity-Time (PT) symmetry so that the detection band will be significantly broadened within the kilo-Hertz range. In this work, we carefully analyze the quantum-noise-limited sensitivity and the dynamical stability of this design. Based on our protocol, our result shows that the quantum-noise-limited sensitivity will be improved by one order of magnitude around 3kHz, which indicates the potential of our design for a future search of neutron star merger signals.
\end{abstract}

\maketitle
\section{Introduction}
Gravitational waves (GWs) radiated from binary neutron star (BNS) inspirals have been detected in many events (e.g. GW170817) by the advanced ground-based laser interferometer GW detectors (LIGO and VIRGO)\,\cite{GW170817,aasi2015advanced,acernese2014advanced,PhysRevD.88.043007,LVC2019GWTC1,LVC2020GWTC2}
. However, the GWs radiated by the BNS post-merger remnants, of which the predicted frequency is around kilo-Hertz\,(kHz), have not been detected yet, due to the limitations of sensitivity at high frequencies\,\cite{Martynov2019,Zhang2021toward}. These high-frequency GWs carry rich physics. For example, They may reveal the details of the center engine of a short gamma-ray burst, the equation of state of ultra-dense nuclear/quark matter, etc. Upgrading the sensitivity of ground-based GW detectors at kilo-Hertz frequency is an important experimental task for the future GW astrophysics\,\cite{Hotokezaka2013,Bauswein2012,Takami2014,Flanagan2008,Kotake2013Multiple,Messenger2012}.

The response of the current LIGO, VIRGO and KAGRA configuration to the high-frequency GWs is limited by the interferometer's bandwidth\,\cite{mizuno1995comparison}. The amplitude of GW-induced-sidebands of the main carrier light decreases with the increase of the GW frequency $\Omega$. The quantum shot noise, which dominates the noise floor at kHz, has a white spectrum\,\cite{Kimble2001Conversion,braginsky1995quantum,braginsky2000energetic,Miao2017Towards,Chen2013Macroscopic,Tsang2011Fundamental}. Various schemes have been proposed for increasing the signal response at high frequency. The simplest way is to increase the intra-cavity power (to e.g.\,10\,MW)\,\cite{Martynov2019}, at the price of sacrificing the low-frequency sensitivity\,(e.g. the proposed new Australia-based instrument NEMO project is targeted on using high power interferometer\,\cite{NEMO}). However, there are many technical challenges to building a high-power Fabry-Perot cavity for GW detection\,\cite{Zhao2005,Zhao2015,Brooks_2021}. Other schemes include: implementing the ``white-light-cavity'' concept to broaden the bandwidth for a detuned interferometer at high frequency; using the resonance created by a long signal recycling cavity; and the design of signal recycling cavity with internal squeezing\,\cite{miao2015enhancing,Miao2018towards,Bentley2019Converting,Adya2020,Thuring:07,PhysRevLett.118.143601,Korobko_2019}.

Recently, a novel scheme was proposed to boost the GW detector sensitivity by reshaping its signal response, in which the interferometer mode $\hat a$ is coupled to a quantum parametric amplifier $\hat c$\,\cite{Li2021enhancing,Li2020Broadband}. In the ideal case, the system Hamiltonian is invariant under the transformation $\hat a\rightarrow \hat c^\dag$ or vice versa, that is, the Hamiltonian has \emph{Parity-Time\,(PT)-symmetry}. In this case, the GW-induced sideband signal fields inside the detector response as $\sim \Omega^{-1}$, which means a large signal boost at low frequency. Compared to the previous white light cavity design\,\cite{miao2015enhancing}, this scheme is dynamically stable since the unstable parametric process will be balanced by the stabilizing sloshing process, and thereby no further feedback control is required for stabilization\,\cite{Li2021enhancing,Gardner2022NondegenerateIS}
. However, the $\Omega^{-1}$ signal response means that most of the advantage is obtained at low frequency, which is easily contaminated by the back-action noise due to the fluctuating quantum radiation pressure force as well as various classical noises.

In this paper, we extend this “PT-symmetry” design concept for boosting the signal response at kilo-Hertz frequency, which could be an alternative approach to increasing the detector sensitivity for detecting GWs emitted by BNS post-merger remnants. In this scheme, the main laser is detuned from the resonance of a signal-recycling laser interferometer, which is coupled to an oppositely detuned quantum parametric amplifier. We will thoroughly analyze the conceptual design of this scheme. Our results show that implementing this PT-symmetry design can significantly boost the sensitivity in a relatively large searching band at high frequency.

The outline of this paper is organized as follows. Section\,\ref{sec:2} gives the basic configurations and the result in the ideal case by a Hamiltonian approach based on the single-mode approximation. Then in Section\,\ref{sec:3} and\,\ref{sec:4}, we perform a detailed analysis on the effect of PT-symmetry breaking to the sensitivity and system dynamical stability. In Section\,\ref{sec:5} , we compute the sensitivity curve using the transfer matrix approach, which is beyond the single-mode and resolved sideband approximations, considering various noise sources. Finally, we give the astrophysical implications of our protocol in Section\,\ref{sec:6}. 

\section{Theoretical principle of the scheme}\label{sec:2}

\begin{figure}[h!]
\centering
\includegraphics[width=0.5\textwidth]{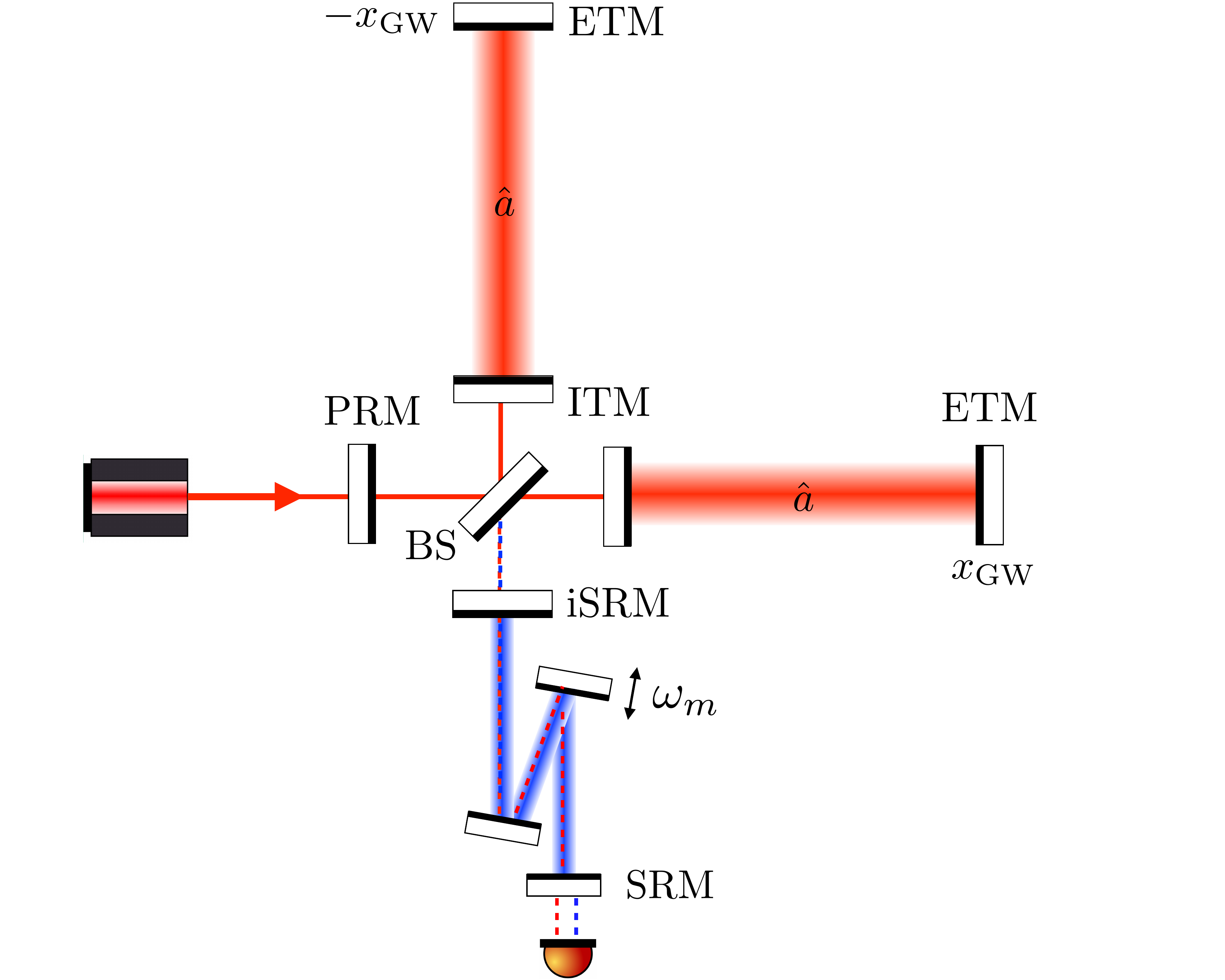}
\caption{Schematic setup: a PT-symmetric Gravitational Wave (GW) Detector. The internal signal recycling mirror (iSRM) at the dark port detunes the main interferometer resonance. An optomechanical device is coupled to the main interferometer which contributes to the parametric process. The red and blue dashed lines are the optical fields at the signal and idler channels. Finally, both of these two channels should be measured to obtain the optimized sensitivity curve.}\label{fig:schematic_setup}
\end{figure}

The basic concept of the scheme can be illustrated (schematically shown in Fig.\,\ref{fig:schematic_setup}) by the following idealized mode interaction Hamiltonian \,\cite{Buonanno2003Scaling,Li2021enhancing}:
\be\label{eq:Hamiltonian_ideal}
\hat H_{\rm int}/\hbar=i\omega_s(\hat a\hat b^\dag e^{i\Delta t}-\hat a^\dag \hat b e^{-i\Delta t})+i G(\hat b^\dag\hat c^\dag e^{i\delta t}-\hat b\hat c e^{-i\delta t}).
\ee
Here the $\hat a$ is the annihilation operator of the differential optical mode of the main signal-recycling interferometer; The $\Delta$ is the detuning of the main laser beam with respect to the $\hat a$ mode, which is introduced by the signal recycling cavity\,\cite{Buonanno2003Scaling}. This non-zero detuning $\Delta$ creates an optical resonance at $\Omega=\Delta$, which improves the signal response around $\Delta$\,\cite{Buonanno2003Scaling}. At the same time, this detuning also leads to the optical spring resonance at low-frequency \,\cite{Buonanno2003Scaling}. Since the high-frequency sensitivity is mainly concerned here, we temporarily ignore the optical spring effect in this section 
for simplicity.
The GW-induced sideband fields are extracted from the $\hat b$ mode, which is parametrically coupled to the $\hat c$ mode. This parametric interaction could have different realizations, for example, using an optomechanical device, or a pumped nonlinear crystal\,\cite{Li2021enhancing,Bently2021}. Unlike the previous work, the frequency matching of the parametric coupling here is generally not perfect, that is, $\delta\neq 0$. 

This Hamiltonian is invariant under the PT-transformation $\hat a e^{i\Delta t}\rightarrow \hat c^\dag e^{i\delta t}$ when the PT-symmetry conditions (incl. $G=\omega_s$ and $\delta=-\Delta$) are satisfied. The Heisenberg equations of motion, considering the coupling between $\hat a$ and the GWs with strength $\alpha$ are:
\be\label{eq:eom}
\begin{split}
&\dot{\hat a}(t)=-i\Delta \hat a(t)-\omega_s\hat b(t)+i\alpha h(t),\\
&\dot{\hat c}^\dag(t)=i\delta \hat c^\dag(t)+G\hat b(t) ,\\
&\dot{\hat b}(t)=-\gamma\hat b(t)+\omega_s\hat a(t)+G\hat c^\dag(t)+\sqrt{2\gamma}\hat b_{\rm in}(t),
\end{split}
\ee
where $\gamma$ is the coupling rate between mode $\hat b(t)$ and external bath $\hat b_{\rm in}(t)$. The outgoing field is $\hat b_{\rm out}(t)=-\hat b_{\rm in}(t)+\sqrt{2\gamma}\hat b(t)$.

\begin{figure}[h!]
\centering
\includegraphics[width=0.5\textwidth]{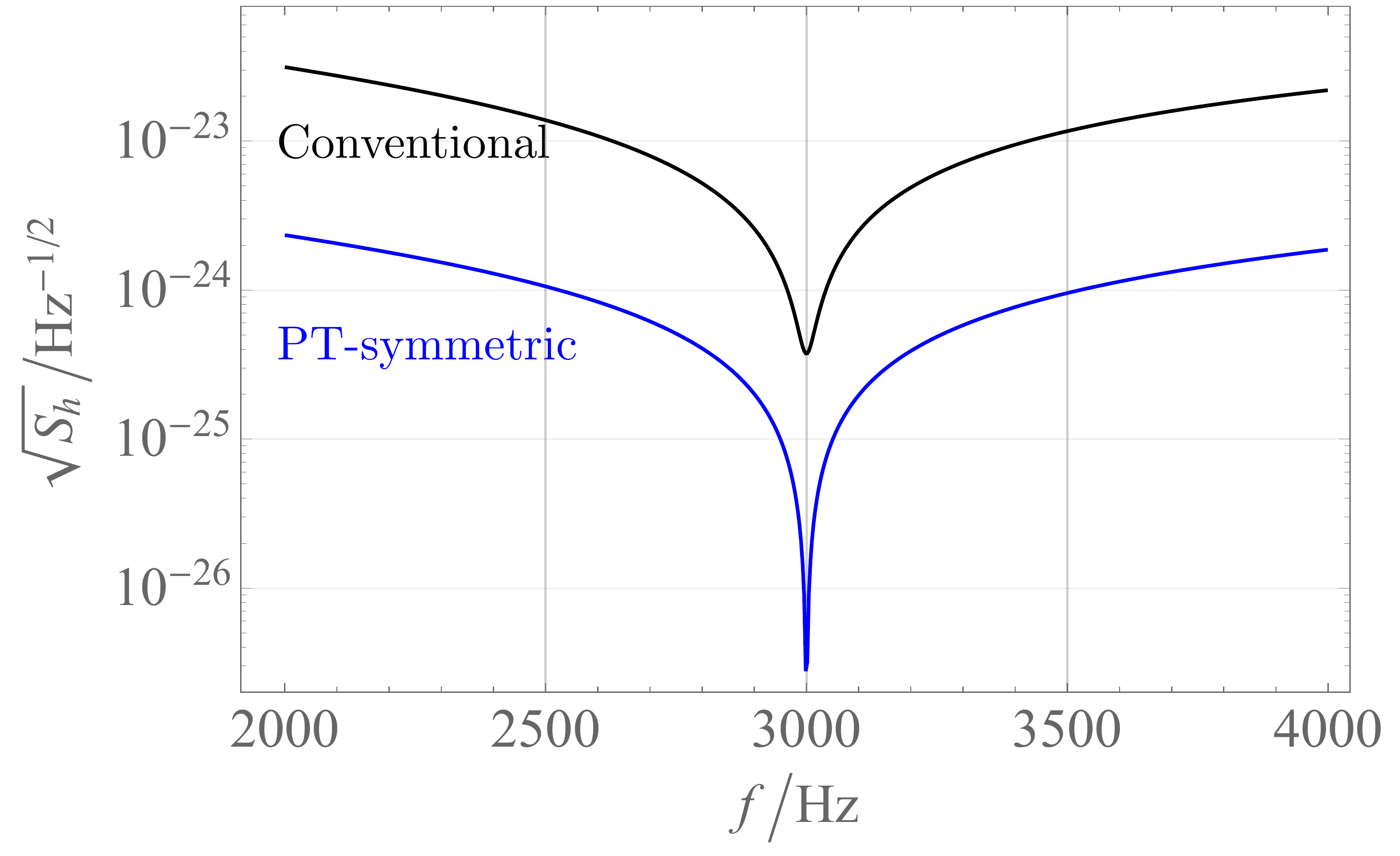}
\caption{The idealized kilo-Hertz GW noise spectrum of gravitational wave detectors enhanced by PT-symmetric configurations, in comparison to the conventional detuned LIGO configuration. The detector dynamics here are described by Eq.\,\eqref{eq:eom}. The parameters we used here is based on mapping the sample parameters in Tab.\,\ref{tab:parameters} of the interferometer to the effective mode-mode interaction model.}\label{fig:ideal_compare}
\end{figure}

Solving these equations of motion under the PT-symmetry condition leads to the input-output relation as:
\be
\begin{split}
&\hat b_{\rm out1}(\Omega)=\frac{\Omega-i\gamma}{\Omega+i\gamma}\hat b_{\rm in1}(\Omega)+\frac{2\alpha\sqrt{\gamma}\omega_s\Delta}{(\Omega+i\gamma)(\Delta^2-\Omega^2)} h(\Omega),\\
&\hat b_{\rm out2}(\Omega)=\frac{\Omega-i\gamma}{\Omega+i\gamma}\hat b_{\rm in2}(\Omega)+\frac{2\alpha\sqrt{\gamma}\omega_s\Omega}{(\Omega+i\gamma)(\Delta^2-\Omega^2)} h(\Omega),
\end{split}
\ee
where we have the amplitude and phase quadrature of optical fields defined in the two-photon formalism\,\cite{Caves1985New} as:
\be
\begin{split}
&\hat b_1(\Omega)=\frac{1}{\sqrt{2}}[\hat b(\Omega)+\hat b^\dag(-\Omega)],\\
&\hat b_2(\Omega)=\frac{1}{\sqrt{2}i}[\hat b(\Omega)-\hat b^\dag(-\Omega)].
\end{split}
\ee
Since we are interested in high frequency region where $\Omega\sim \Delta$,  the signal response of $\hat b_{\rm out1}$ and $\hat b_{\rm out2}$ are roughly the same and  scales approximately as $\sim 1/(\Delta-\Omega)$.

The shot-noise-limited sensitivity of the detuned PT-symmetric scheme, quantified by the signal referred shot noise spectral density $S_{hh}(\Omega)$, is given by (suppose the phase quadrature $\hat b_{\rm out2}$ is measured):
\be
S^{\rm PT}_{hh}(\Omega)\approx\frac{(\Delta^2-\Omega^2)^2(\Omega^2+\gamma^2)}{4\gamma\omega_s^2\alpha^2\Omega^2},
\ee
while for the conventional detuned interferometer (also the phase quadrature is measured) given by:
\be
S^{\rm con}_{hh}(\Omega)=\frac{(\Omega^2-\gamma^2-\Delta^2)^2+4\gamma^2\Omega^2}{4\alpha^2\gamma(\Omega^2+\gamma^2)}.
\ee
The comparison between the sensitivities of these two configurations at $\Omega\sim \Delta$ is shown in Fig.\,\ref{fig:ideal_compare}. At the resonance point $\Omega=\Delta$, the PT-symmetry scheme has a larger boost due to its response $\sim 1/(\Omega-\Delta)$ in the ideal case. 

In this ideal case, the system is also dynamically stable which can be understood from the following analysis. Adiabatic elimination of the $\hat b-$field\,(usually has a large bandwidth) leads to  the equations of motion for $\hat a,\hat c$:
\be
\begin{split}
&\left(\frac{d}{dt}+i\Delta+\frac{\omega_s^2}{\gamma}\right)\hat a=-\frac{\omega_s G}{\gamma}\hat c^\dag-\sqrt{\frac{\omega_s^2}{2\gamma}}\hat b_{\rm in}+i\alpha h,\\
&\left(\frac{d}{dt}-i\delta-\frac{G^2}{\gamma}\right)\hat c^\dag=\frac{\omega_s G}{\gamma}\hat a+\sqrt{\frac{G^2}{2\gamma}}\hat b_{\rm in}.
\end{split}
\ee
There is a damping factor $\omega_s^2/\gamma$ for the $\hat a$ field while an anti-damping factor $-G^2/\gamma$ for the $\hat c^\dag$ field. Under the PT-symmetric conditions $G=\omega_s$ and $\delta=-\Delta$, the anti-damping factor and damping factor cancel each other for the effective 
mode $\hat a+\hat c^\dag$:
\be\label{eq:a+c}
\left(\frac{d}{dt}+i\Delta\right)(\hat a+\hat c^\dag)=i\alpha h.
\ee


Note that in Fig.\,\ref{fig:ideal_compare}, there is one peak at $\Omega=\Delta$, while we actually have three different modes $\hat a,\hat b,\hat c$ in our Hamiltonian, which means there exists degeneracy due to the PT symmetry\,\cite{Wiersig2016,Zhang2019}. This degeneracy can be easily understood since $\hat a$ is equivalent to $\hat c^\dag$, and the $\hat b$-field couples to the $\hat a,\hat c$ fields in such a way that there is no effect on the resonance frequency of the $\hat b$-mode (see Eq.\,\eqref{eq:a+c}). Practical imperfections of our system will cause the breaking of the PT-symmetry, which will affect the detector sensitivity and the system stability.

This degeneracy can be understood from the algebraic structure of the Heisenberg equations of motion Eq.\,\eqref{eq:eom}.  If we take the representation which chooses $\hat a,\hat b,\hat c^\dag$ to be the basis vectors, the Heisenberg equation of motion becomes:
\be
\label{eq:eom_matrix}
\frac{d}{dt}\hat{\mathbf v}= \hat{\mathcal{D}} \hat{\mathbf v} + \mathbf s(t),
\ee
where $\hat{\mathbf v}\equiv( v_1, v_2, v_3)^T$ is a combined mode consisting $v_1,v_2,v_3$ of $\hat a,\hat b,\hat c^\dag$,respectively; 
$\mathbf s(t)$ is the vector describes the signal and noise adding to the system:
\be
\mathbf s(t)=(i\alpha h(t),\sqrt{2\gamma}\hat{b}_{in}(t),0)^T;
\ee
$\hat{\mathcal{D}}$ is the dynamic matrix:
\be
\hat{\mathcal{D}}=
\left[
\begin{array}{ccc}
-i\Delta&-\omega_s&0\\
\omega_s&-\gamma&G\\
0&G& i\delta
\end{array}
\right].
\ee
The eigenvalues $\lambda_{1,2,3}$ and the corresponding eigenvectors $\hat{\mathbf v}_{1,2,3}$ then describes the poles and its corresponding modes of the whole system, respectively. The response in the frequency domain can be written as:
\be
\mathbf v(\Omega) = -\frac{\mathbf s(\Omega)}{\lambda+i\Omega}.
\ee

\begin{figure}[h!]
\centering
\includegraphics[width=0.48\textwidth]{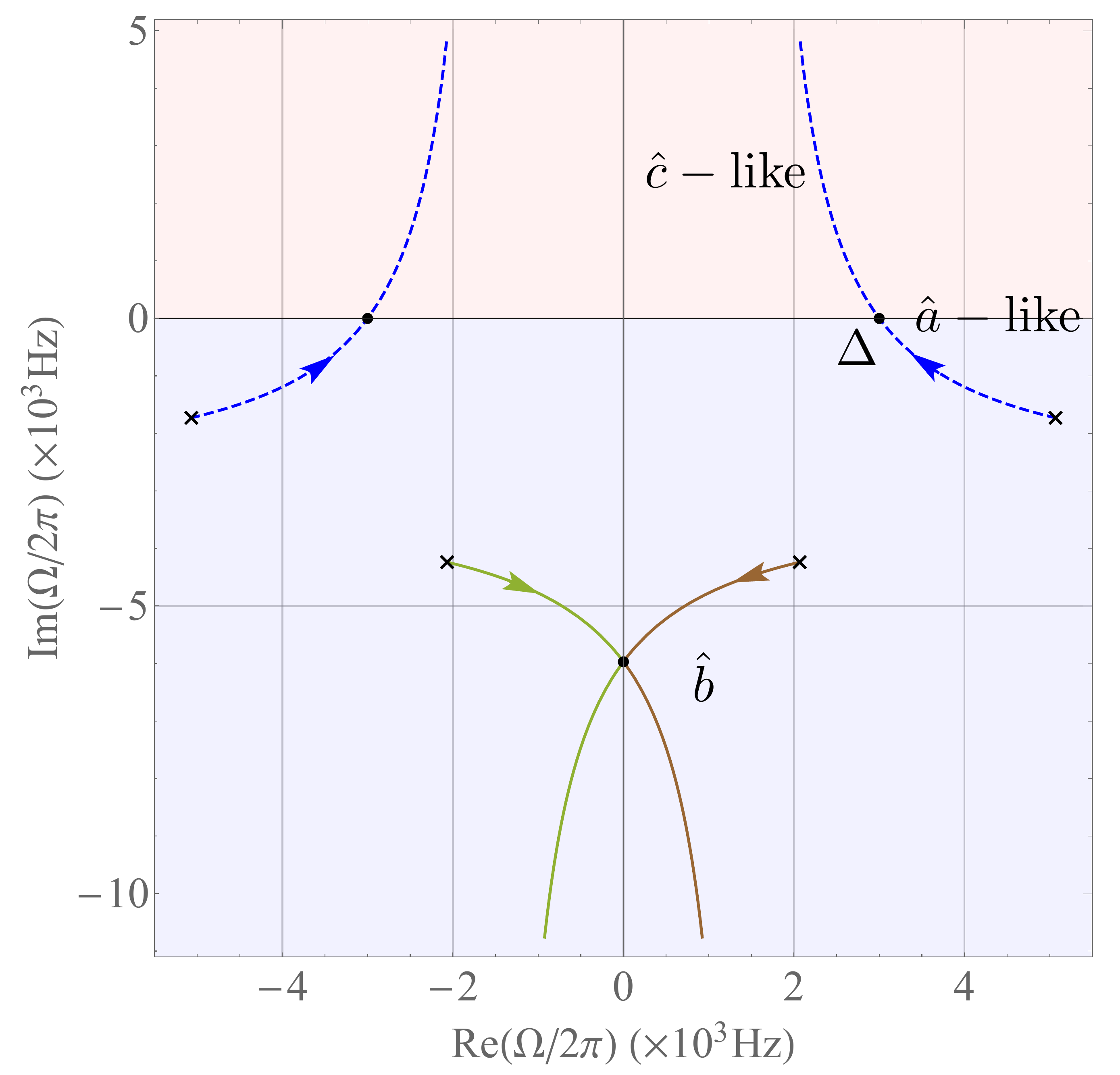}
\caption{The trajectories of the system's eigenvalues based on the ideal Hamiltonian with varies $G$ form 0 to 2$\omega_s$ and fixed $\omega_s$. Note that these trajectories are symmetric on the left and right plane since the Hamiltonian is Hermitian. As discussed in the main text, one of the eigenmodes has a fixed eigenvalue $\lambda_1=-i\Delta$, which corresponds to a fixed pole at frequency $\Omega_1=-i\lambda_1=\Delta$. The other two eigenvalues\,(poles) change with the variation of $G$, they are shown as the blue dashed lines\,($\lambda_3$) and the solid lines\,($\lambda_2$). Under the PT-symmetry condition $G=\omega_s$, the poles of $\lambda_1$ and $\lambda_3$ have a degeneracy. Then if $G$ varies across $\omega_s$, the pole of $\lambda_3$ gets a positive imaginary part and becomes unstable. Throughout this variation, $\mathbf v_3$ is mainly consist of $\hat a$\, ($v_1=-\omega_s/G$) and $\hat c^\dag$\,($v_3=1$). Therefore we denote the mode "$\hat c$-like" field after crossing the degeneracy point since $\omega_s<G$, while it's denoted as "$\hat{a}$-like" field before since $\omega_s>G$.
}
\label{fig:root_symmetry_breaking}
\end{figure}

Satisfying one of the PT-symmetry conditions $\delta=-\Delta$, the eigenvalues and the corresponding eigenvectors are given by:
\be
\begin{split}
&\lambda_{1}:\quad \mathbf v_1=(-\frac{G}{\omega_s},0,1)^T,\\
&\lambda_{2}:\quad \mathbf v_2=(-\frac{\omega_s}{G},-\frac{\gamma-i\Delta}{2\chi}-\frac{\lambda_3-\lambda_2}{2\chi},1)^T,\\
&\lambda_{3}:\quad \mathbf v_3=(-\frac{\omega_s}{G},-\frac{\gamma-i\Delta}{2\chi}+\frac{\lambda_3-\lambda_2}{2\chi},1)^T,\\
\end{split}
\ee
where 
\be
\begin{split}
&\lambda_1=-i\Delta,\\
&\lambda_{2,3}=\frac{1}{2}[-\gamma-i\Delta\mp\sqrt{(\gamma-i\Delta)^2+4(G^2-\omega_s^2)}].
\end{split}
\ee
The other PT-symmetry condition $G=\omega_s$ would lead to a degeneracy since $\mathbf{v}_{1/3}$ corresponds to the same eigenvalue $\lambda_1=\lambda_3=-i\Delta$. Breaking this PT-symmetry condition would lead to the breaking of this degeneracy, as shown in Fig.\,\ref{fig:root_symmetry_breaking} where the eigenvalues are plotted in the complex frequency domain. This figure also shows that there will be an unstable mode in some parameter regions, of which the eigenvalue is located on the upper complex plane. This stability issue will be further explored in Section\,\ref{sec:4}.

\section{Effect of PT-symmetry breaking: sensitivity}\label{sec:3}
\begin{figure}[h!]
\centering
\includegraphics[width=0.4\textwidth]{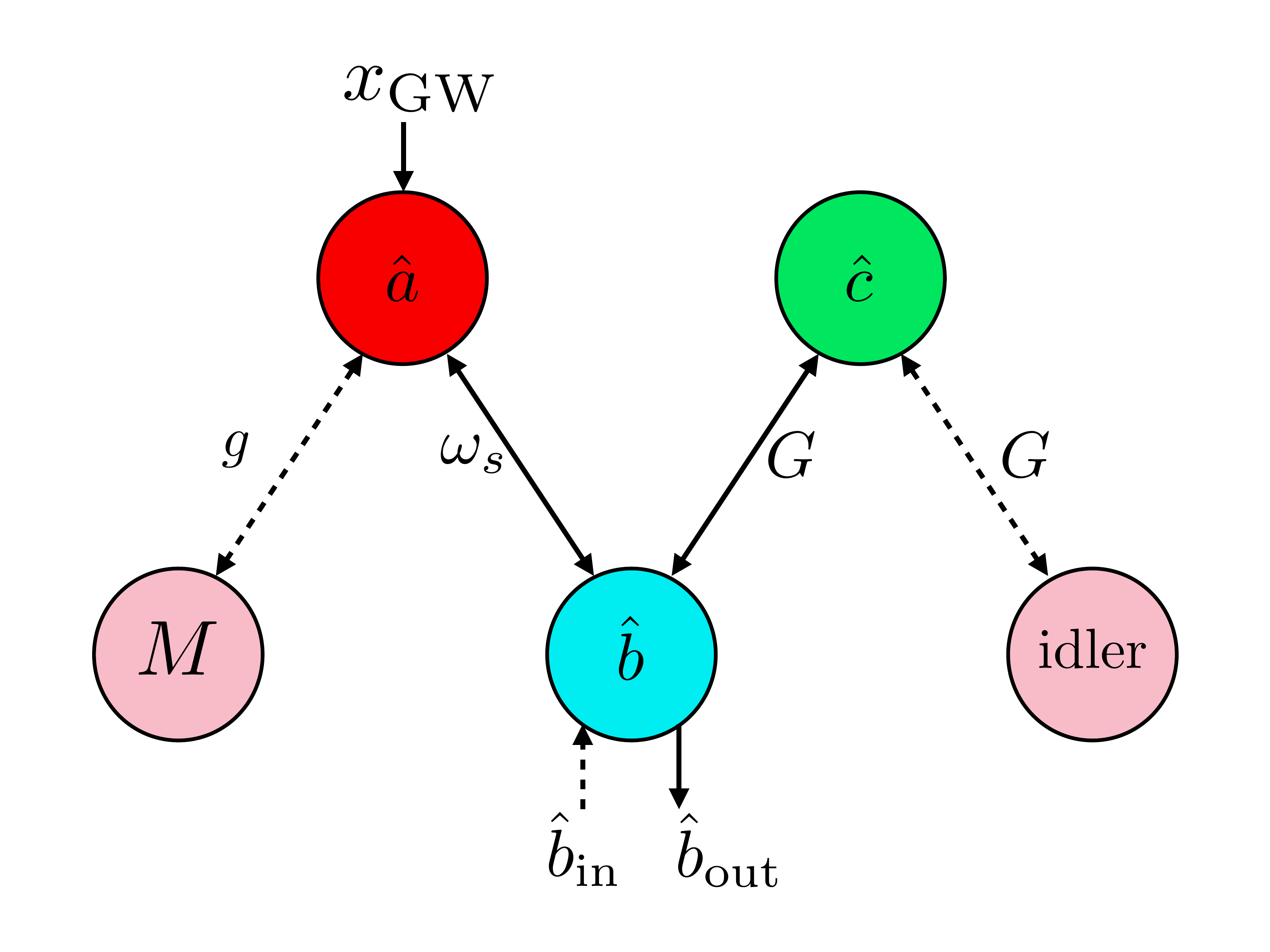}
\caption{The breaking of the PT-symmetry due to practical system imperfections: the introduction of the off-resonant sideband fields inside the optomechanical filter cavity\,(idler channel) and the pondermotive interactions in the interferometer arm cavities. The test mass and the idler field do not couple to the $\hat a$ and $\hat c$ modes in a symmetric way, thereby breaking the PT-symmetry of the entire system.}\label{fig:symmetry_breaking}
\end{figure}

The main practical factors that lead to the PT-symmetry breaking can be generally summarized as follows (also see Fig.\,\ref{fig:symmetry_breaking}): (1) the off-resonance sidebands in the optomechanical filter cavity, which we ignored in the ideal case in Eq\,\eqref{eq:Hamiltonian_ideal}; (2) the pondermotive interaction happens inside the interferometer arm cavities\,\cite{Buonanno2003Scaling}. In this section, we focus on their influences on the sensitivity of our protocol.

\subsection{Influence from the off-resonant sidebands}
\begin{figure}[h!]
\centering
\includegraphics[width=0.5\textwidth]{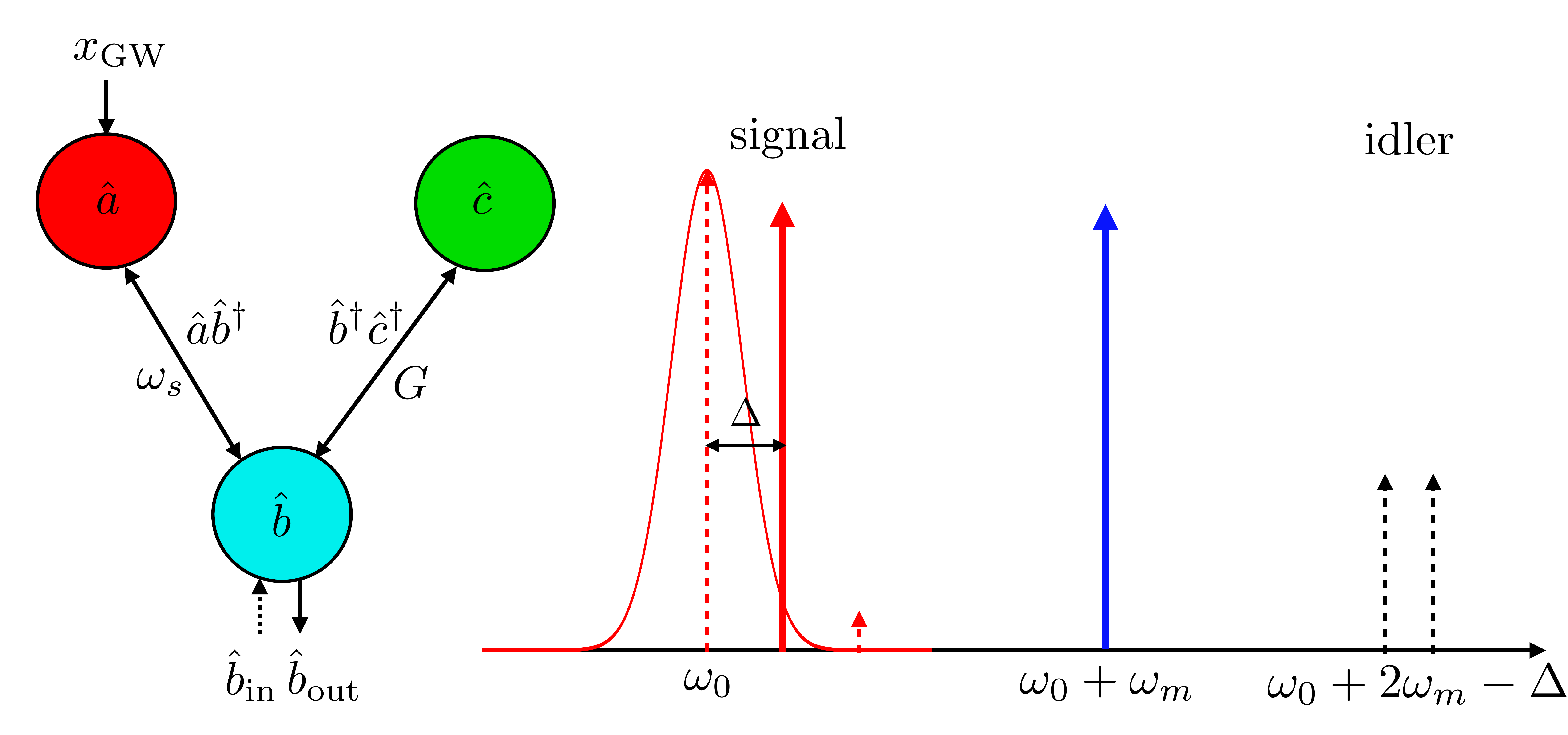}
\caption{Left panel: the structure of the mode-coupling Hamiltonian, where the $\hat a$ and $\hat c$ are the main interferometer mode and the mechanical mode, which are PT-symmetric to each other. For detecting the high-frequency GWs, detuning should be introduced to these two modes. Right panel: the sideband fields on-resonance with $\omega_0$ are defined as \emph{signal channel}, while the far off-resonance sidebands resonance at $\sim\omega_0+2\omega_m$ is defined as \emph{idler channel}. The red curve sketches the main interferometer  resonance profile.}\label{fig:sidebands}
\end{figure}

A typical realization of the parametric interaction is to use an optomechanical device with the interaction Hamiltonian $\hat H_{\rm om}=\hbar \tilde{G} \hat b^\dag\hat b\hat x_m$, where $\tilde{G}=\omega_0/L_b$ is the single-photon optomechanical coupling strength and $L_b$ is the length of the filter cavity, the $\hat x_m$ is the displacement operator of the mechanical oscillator. Supposing the system is pumped with detuning $\omega_m$, the optomechanical interaction can be written as:
\be
\begin{split}
\hat H_{\rm om}/\hbar= G  &\left(\hat b\hat c e^{i\delta t}+
\hat b^\dag\hat c^\dag e^{-i\delta t}+\right.\\
&\left.\hat b^\dag\hat c e^{-i(2\omega_m-\delta) t}+\hat b\hat c^{\dag} e^{i(2\omega_m-\delta) t}\right),
\end{split}
\ee
where $G=\tilde{G}x_{\rm zpf}\bar b$ with $x_{\rm zpf}$ the zero-point displacement of the mechanical oscillator and $\bar b$ the mean amplitude of the pumping beam. In the previous section, we ignored the last two terms of the above Hamiltonian using the resolved sideband limit to obtain Eq.\,\eqref{eq:Hamiltonian_ideal}. However, in reality, these far-off resonant sidebands will have a non-negligible degradation to the detector sensitivity. In this work, we follow the standard quantum optics terminology and name the optical sideband fields around $\omega_0$ to be the ``\emph{signal channel}" and those around $\omega_0+2\omega_m$ to be the ''\emph{idler channel}'', as shown in Fig.\,\ref{fig:sidebands}. For illustrative purposes, this subsection will study this effect, while temporarily disregarding the pondermotive effect in analyzing the idler fields.

\begin{figure}[h!]
\centering
\includegraphics[width=0.5\textwidth]{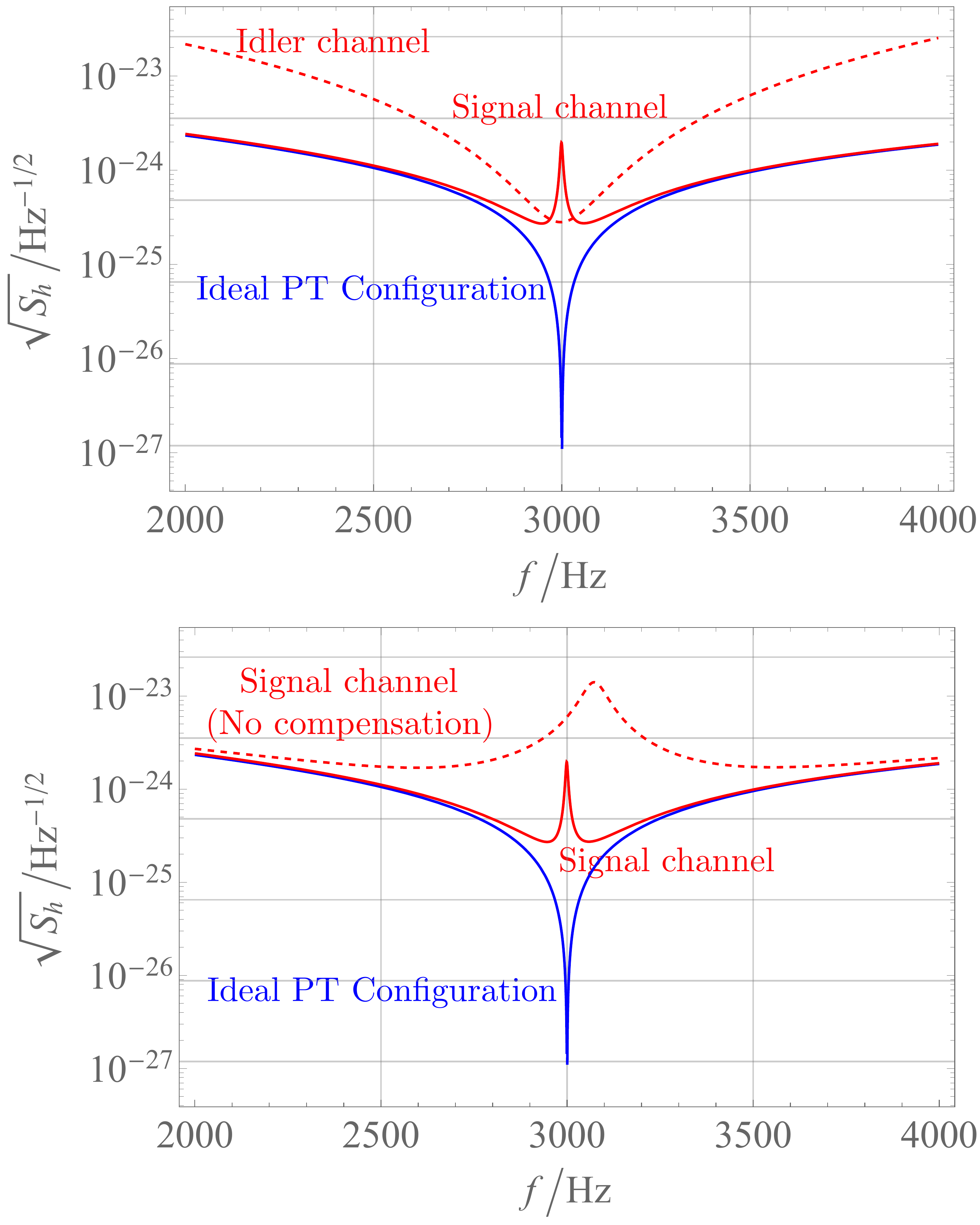}
\caption{The detector sensitivity considers purely the effect of idler fields (i.e. the pondermotive effect is ignored). Upper panel: the sensitivity of both the idler and signal channels, taking into account the frequency compensation $\omega_{\rm opt}$. Lower panel: The sensitivity of the signal channel with/without the frequency compensation. The two dips around the detuning frequency is a manifestation of degeneracy breaking introduced by the idler channel.
}\label{fig:idler_effect}
\end{figure}

Introducing the idler channel around $\omega_0+2\Omega_m$ will break the PT-symmetry as shown in Fig.\,\ref{fig:idler_effect}, and bring the following two effects.

$\bullet$ \textbf{Optical spring effect}. The optical spring effect in the optomechanical filter cavity will modify the mechanical frequency to be $\omega^{\rm opt}_m\approx\omega_m+\omega_{\rm opt}$, where  $\omega_{\rm opt}=cP_b\omega_0/2m\omega_m^2c^2L_b$. This optical spring effect must be compensated, otherwise, the PT-symmetry will be broken and the detector will not be able to reach its optimal sensitivity.  This effect is plotted in the lower panel of Fig.\,\ref{fig:idler_effect}, which demonstrates the importance of this frequency compensation. 

$\bullet$ \textbf{Correction to the GW signal response.} 
The GW response of the detector will be modified in two ways by introducing the idler channel. First, the mode degeneracy introduced by PT-symmetry will be broken. This splits the single-peak at $\Omega=\Delta$ in the sensitivity curve as shown in Fig.\,\ref{fig:idler_effect}.
Second, the GW signal sidebands will flow into the idler channels. Therefore GW information can also be extracted from the detection of the optical quadratures at the idler channels:
\be
\begin{split}
&\hat b_{i1}(\Omega)=\frac{1}{\sqrt{2}}[\hat b(2\omega_m+\Omega)+\hat b^\dag(2\omega_m-\Omega)],\\
&\hat b_{i2}(\Omega)=\frac{1}{\sqrt{2}i}[\hat b(2\omega_m+\Omega)-\hat b^\dag(2\omega_m-\Omega)].\\
\end{split}
\ee
Since the idler channel is separated far from the signal channel in the frequency domain,  this means we can use two homodyne detectors to measure both the signal channel and idler channel, and then do data-post-processing to optimize the detector sensitivity, which we will explain in detail later. This analysis also shows that a full transfer matrix analysis is needed for a more accurate sensitivity curve.

\subsection{Pondermotive effect}
Now we discuss the pondermotive effect which has been ignored in the previous sections. For a detuned main interferometer, the pondermotive interaction generates an additional stiffness for the test masses\,\cite{Buonanno_Chen_2001}, which is the optical spring effect\,\cite{Buonanno2003Scaling}. This optical stiffness comes from the dependence of radiation pressure force on the test mass displacement. Adding the pondermotive effect term $+ig\hat X$ to the right-hand side of Eq.\,\eqref{eq:eom} ($g=\omega_0\alpha/L$ is the linearised optomechanical coupling constant in the arm cavity, $X$ is displacement of test mass.) leads to:
\be
\dot{\hat X}=\hat P/M,\quad
\dot{\hat P}=\hbar g(\hat a+\hat a^\dag).
\ee 
The optical spring rigidity $K_{\rm opt}(\Omega)$ under the PT-symmetry condition can be solved as (where the PT-symmetry conditions $G=\omega_s$ and $\delta=-\Delta$ have been used):
\be\label{eq:kopt}
K_{\rm opt}(\Omega)=2\hbar g^2\left[\frac{\Delta}{\Delta^2-\Omega^2}-\frac{2\Omega\Delta G^2}{(\Omega+i\gamma)(\Delta^2-\Omega^2)^2}\right].
\ee
In the following, it will be written as $K_{\rm opt}(\Omega)=K_{\rm opt1}(\Omega)+K_{\rm opt2}(\Omega)$, where the $K_{\rm opt1}(\Omega)=2\hbar g^2\Delta /(\Delta^2-\Omega^2)$, which only depends on the optomechanical coupling $g$ in the arm cavity. This $K_{\rm opt1}(\Omega)$ resembles the rigidity of a detuned \emph{perfect cavity with zero bandwidth}, which reflects the fact that the loss and gain are balanced under the PT-symmetry condition.

The optical rigidity will become very large at $\Omega=\Delta$ due to the significantly boosted displacement-induced-sideband fields as $\propto \alpha \R{X}(\Omega)/(\Delta^2-\Omega^2)$. In other words, at $\Omega=\Delta$, the test masses will become so stiff that the external force can not drive their motions. Therefore the signal at this frequency is strongly suppressed and a peak in the sensitivity curve is expected. 

The input-output relations considering the pondermotive effect can be written as:
\be
\mathbf{\hat b}_{\rm out}(\Omega)
=e^{i\beta(\Omega)}\mathbb{M}(\Omega).\mathbf{\hat b}_{\rm in}(\Omega)
+\mathbf{D}(\Omega)h(\Omega),
\ee
in which we have: $e^{i\beta(\Omega)}=(\Omega-i\gamma)/(\Omega+i\gamma)$
and
\be
\begin{split}
&\mathbb{M}(\Omega)=\\&\chi_M(\Omega)
\left[
\begin{array}{cc}
-\chi^{-1}_M(\Omega)+K^{i}_{\rm opt2}(\Omega)&-(\Delta/\Omega)K_{\rm opt2}^{r}(\Omega)\\
-(\gamma/\Delta)K^i_{\rm opt2}(\Omega)&-\chi^{-1}_M(\Omega)+K^{i}_{\rm opt2}(\Omega)
\end{array}
\right],
\end{split}
\ee
where $\chi_M(\Omega)$ is the test mass mechancial response function modified by the pondermotive effect:
\be
\chi_M(\Omega)=-\frac{1}{M\Omega^2-K_{\rm opt}(\Omega)}.
\ee
The signal response matrix is:
\be
\mathbf{D}(\Omega)=-\frac{2ig\sqrt{\gamma}G}{(\Omega+i\gamma)(\Delta^2-\Omega^2)}\chi_M(\Omega)M\Omega^2L
\left[
\begin{array}{c}
\Delta\\
\Omega
\end{array}
\right],
\ee
where at $\Omega=\Delta$, the signal response vanishes, and the sensitivity will have a very sharp peak as expected. This effect will not happen for the tuned PT-symmetric interferometer\,\cite{Li2021enhancing} since there is no optical spring effect when $\Delta=0$. The problem of sensitivity degradation due to this pondermotive effect can be solved using (1) an optimal combination of different measurement channels which is discussed in detail in Sec.\,\ref{sec:5}, and (2) negative inertia.

\begin{figure}[h!]
\centering
\includegraphics[width=0.5\textwidth]{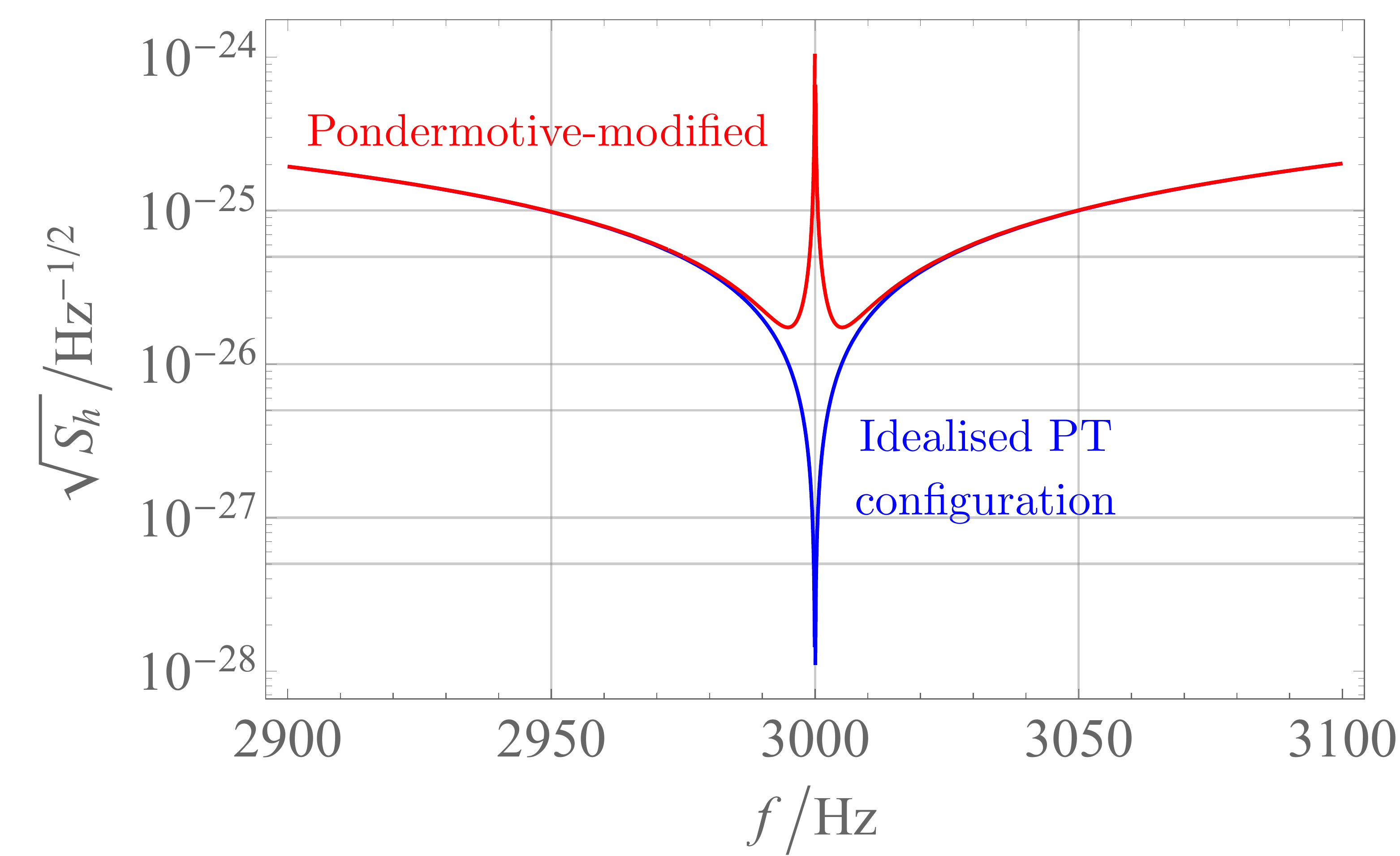}
\caption{The degradation of detector sensitivity is due to the pondermotive interaction between the test masses and the optical fields in the main interferometer. The narrow sensitivity peak at the optical resonance frequency $3000$\,Hz with bandwidth $\sim 10$\,Hz comes from the optical stiffness enhanced by the PT-symmetric configuration. For highlighting the pondermotive effect, the idler channel effect is ignored here.}\label{fig:pondermotive_effect}
\end{figure}


\subsection{Cancelling the pondermotive effect using negative inertia}
The pondermotive effect reflects the fact that the system is not fully PT-symmetric when the test masses dynamics are considered. For achieving a fully PT-symmetric system, a negative mass term\,\cite{Khalili2018,Zeuthen2019,Tsang2010,Tsang2012,PhysRevD.78.062003,PhysRevD.83.062003} 
needs to be introduced with an additional equation of motion:
\be
\dot{\hat x}=-\frac{\hat p}{\mu},\quad \dot{\hat p}=\hbar \tilde{g}(\hat c+\hat c^\dag),
\ee
where $\mu=-M$ and $\tilde{g}=g$ should be satisfied to achieve the PT-symmetry. In this case, the input-output relation becomes very simple:
\be
\left[
\begin{array}{c}
\hat b_{\rm out1}(\Omega)\\
\hat b_{\rm out2}(\Omega)
\end{array}
\right]=e^{i\beta (\Omega)}
\left[
\begin{array}{c}
\hat b_{\rm in1}(\Omega)\\
\hat b_{\rm in2}(\Omega)
\end{array}
\right]+\mathbf{D}_{\rm sym}(\Omega)h(\Omega),
\ee
where we define:
\be
\mathbf{D}_{\rm sym}(\Omega)=\frac{2ig\sqrt{\gamma}\chi}{(\gamma-i\Omega)(\Delta^2-\Omega^2)}\frac{M\Omega^2 L}{-M\Omega^2+K_{\rm opt1}(\Omega)}
\left[
\begin{array}{c}
\Delta\\
-i\Omega
\end{array}
\right].
\ee
At $\Omega=\Delta$, the signal response is finite. The physical realization of negative inertia coupled to the mode $\hat c$  could be realized by optical systems, we leave the detailed design to the accompanying paper. In this paper, we merely introduce this concept without discussing its details. The main approach to cancel the sensitivity peak at $\Omega=\Delta$ discussed in this paper is the optimal combination of the different measurement channels which will be discussed in the next section.

\section{Effect of PT-symmetry breaking: Stability}\label{sec:4}

Instability arise when there are external energy sources pump the system continuously. As shown in Fig.\ref{fig:schematic_setup}, the arm cavity and filter cavity are both pumped in the blue-detuned way. A single optomechanical system pumped by a blue-detuned coherent field would have instability since the Stokes sideband would overwhelm the anti-Stokes sideband so that the pumping light is transferring energy to the mechanical degree of freedom\,\cite{Chen2013Macroscopic,Aspelmeyer2014}.  In our design protocol, the blue-detuned main interferometer is coupled to the blue-detuned filter cavity with the sloshing frequency $\omega_s$, therefore this coupling will certainly affect the system dynamics and stability. In Fig.\,\ref{fig:root_symmetry_breaking}, we have seen that there could be unstable modes when PT-symmetry is broken.

\begin{figure}[h!]
	\centering
		\subfigure[]{
		\includegraphics[width=1.\linewidth]{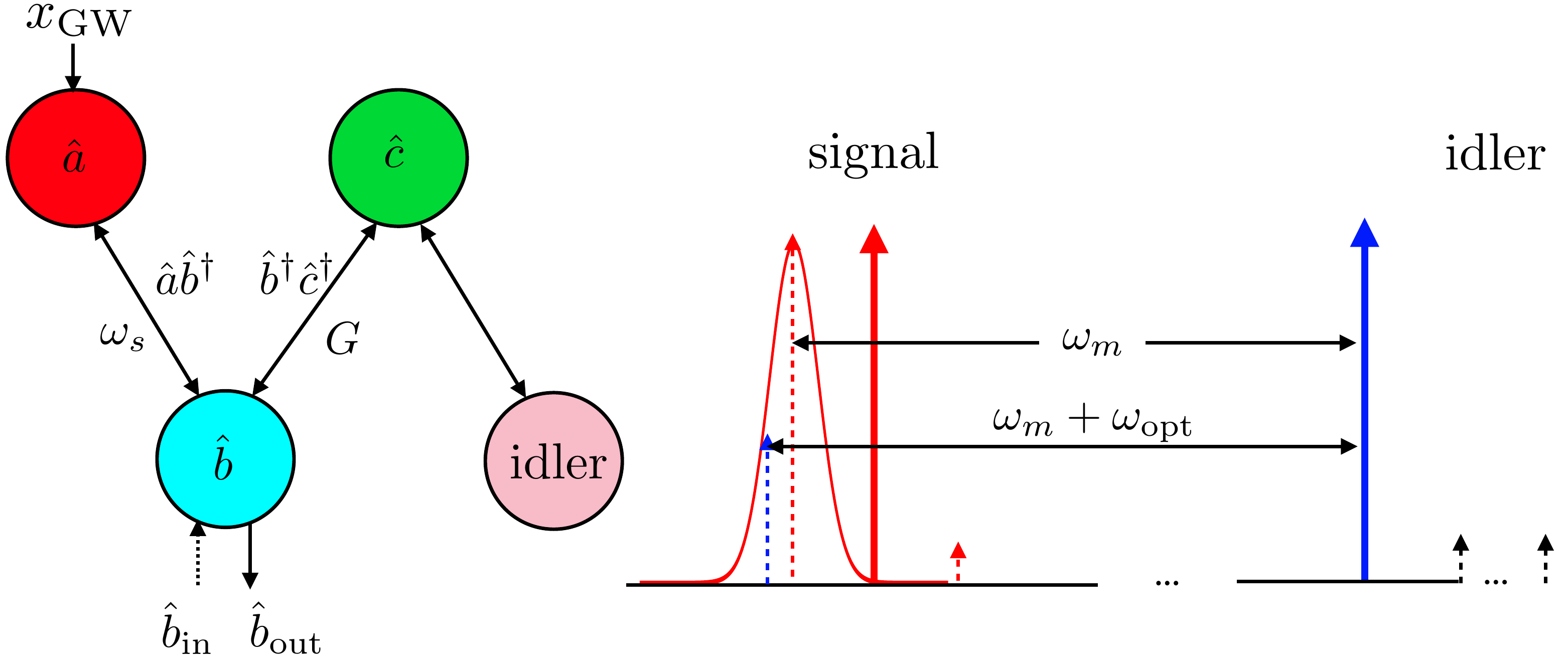}}
	\subfigure[]{
		\includegraphics[width=1.\linewidth]{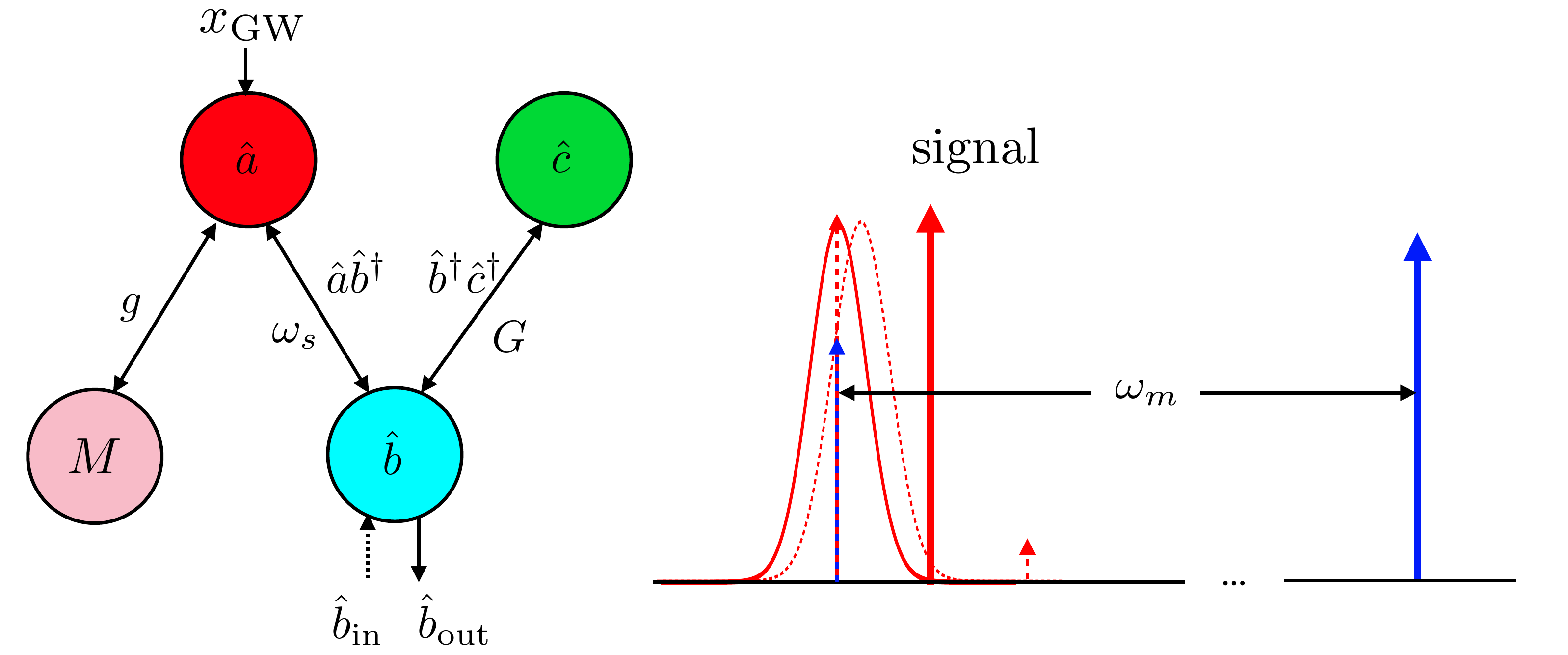}}
	\caption{Instabilities induced by the uncompensated resonant frequency drift introduced by the PT-symmetry breaking. (a) The Stokes sideband of the pumping light in the filter cavity is a bit off-resonant to the main cavity resonance due to the optical spring effect $\omega_{\rm opt}$, which breaks the balance between the parametric and the sloshing process. (b) The Stokes sideband of the pumping light in the filter cavity is also an off-resonant due to the small shift of the main cavity resonance due to its coupling with the test mass, which also breaks the parametric-sloshing balance and creates instability.} \label{fig:instability}
\end{figure}

\subsection{Instability induced by pondermotive effect}

\begin{figure}[h]
    \centering
    \includegraphics[width=0.45\textwidth]{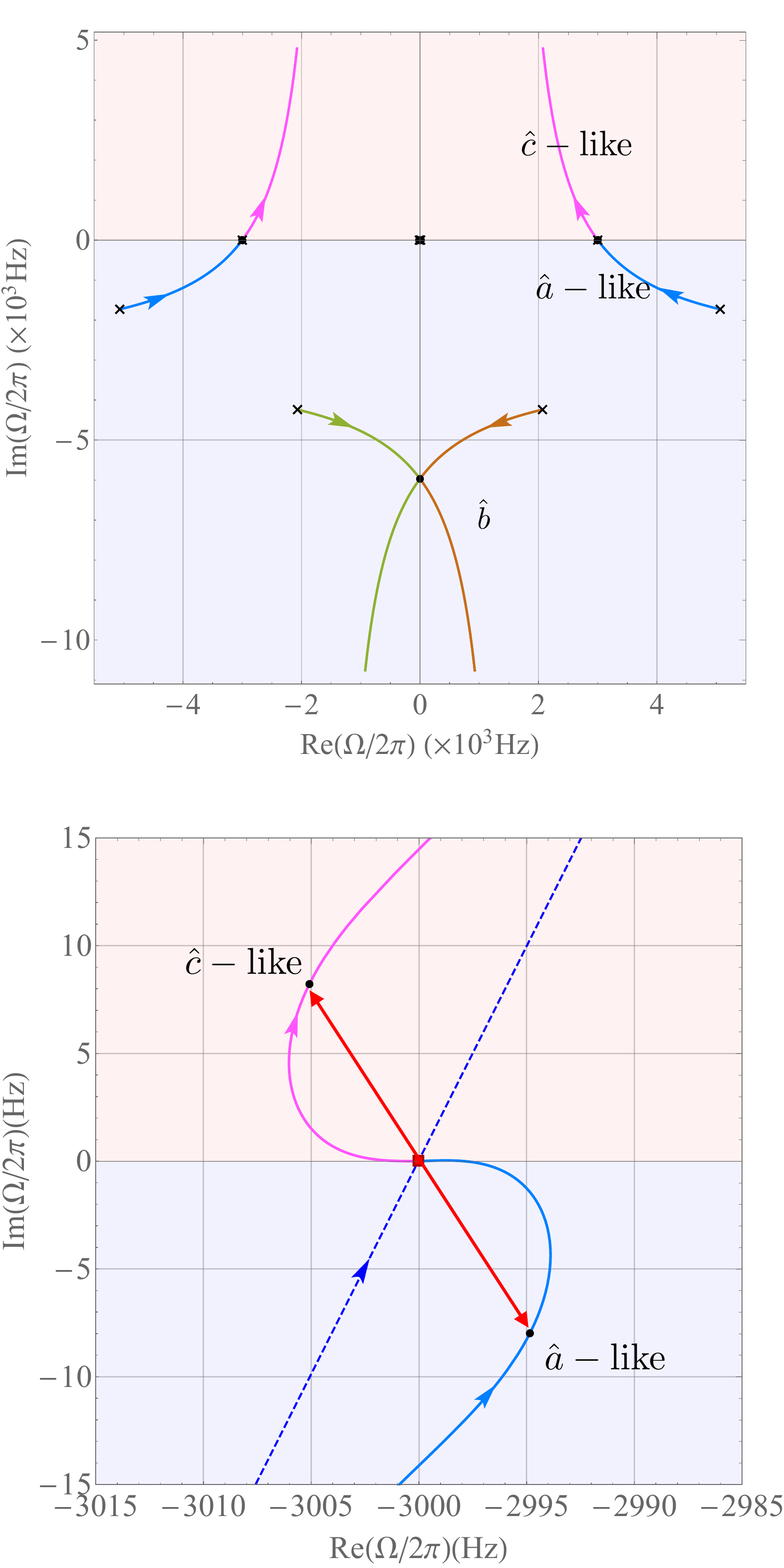}
    \caption{The influence of the pondermotive interaction inside the arm cavities on the resonance pole at high frequencies. Upper panel: the global pole trajectories plotted in the same way as in Fig.\,\ref{fig:root_symmetry_breaking}. Lower panel: zooming-in the pole trajectories near the detuning point $\Omega=\Delta$, where the degenerate root (when $\omega_s=G$) splits into two different poles and introduce instability. This instability can be removed by tuning the pumping field frequency in the filter cavity. The dashed line is the pole trajectory of the ideal case for comparison.}\label{fig:instability_pondermotive}
\end{figure}

Introducing the pondermotive effect in the arm cavity means the coupling between the mechanical motion and the optical mode of the main interferometer, which will shift the optical resonance of the main interferometer\,\cite{Buonanno2003Scaling}. If this frequency shift was left uncompensated,  it will create a suppression of the Stokes sidebands of the pumping field in the filter cavity when being sloshed to the main interferometer\,(see Fig.\,\ref{fig:instability}). Therefore, the balance between the sloshing process and the parametric process is broken and instability forms. In Fig.\,\ref{fig:instability_pondermotive}, we plot the influence of the pondermotive effect on the trajectories of poles, where the trajectories detour near the detuning frequency $\Omega=\Delta$. The degenerate point at $\Omega=\Delta$ under the PT-symmetric condition splits into two different points, one of which is unstable. Such instability can be removed by carefully tuning the frequency of the pumping field.

Besides, the pondermotive effect in the arm cavity also generates a dynamical back-action to the test mass mirrors, which is modified by the coupling to the auxiliary system as shown in Eq.\,\eqref{eq:kopt}. Using the parameter listed in Tab.\ref{tab:parameters}, these modified dynamics contribute a finite optical spring frequency around $7.7$\,Hz with an anti-damping factor equal to $2\pi\times0.019$\,rad/s to the test mass. This $7.7$\,Hz optical spring resonance manifests as a dip around $7.7$\,Hz in the sensitivity curve Fig.\,\ref{fig:global_wiener}. The anti-damping rate $2\pi\times0.019$\,rad/s is well within the LIGO control band. \\

\subsection{Instability induced by the off-resonant sidebands}

After removing the instability near $\Omega=\Delta$ induced by the arm cavity pondermotive effect by tuning the pumping field frequency, the instability introduced by the idler channel (i.e. the off-resonant sideband fields in the filter cavity) also needs to be removed. This instability exists because the introducing of these off-resonant idler fields in the filter cavity brings an optical spring that modifies the mechanical resonance frequency. The mechanism is shown in Fig.\,\ref{fig:instability}. The Stokes sidebands of the pumping field inside the filter cavity are off-resonant with the main interferometer optical resonance,  thereby being suppressed when it is sloshed from the filter cavity to the main interferometer. However, the large bandwidth of the filter cavity indicates that such a small mechanical resonance shift does not affect the parametric process. This means that the parametric process will overwhelm the sloshing process and instability forms. Using the Hamiltonian approach, we can quantitatively compute the effect of the frequency compensation on the stability, the result shows that we need to carefully tune the compensation frequency near the $\omega_{\rm opt}$ to stabilise the system. 

\begin{figure}[h!]
    \centering
    \includegraphics[width=0.48\textwidth]{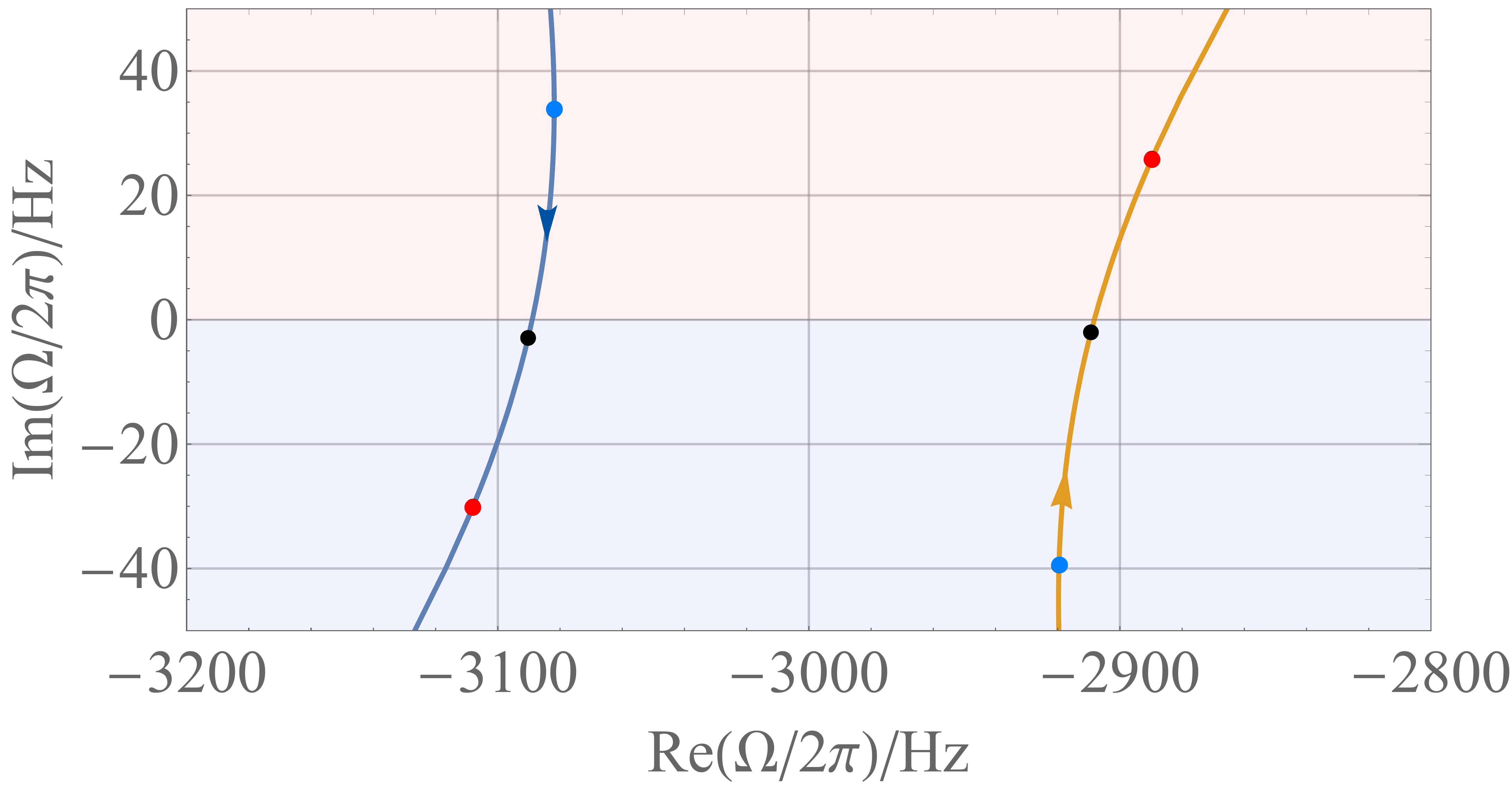}
    \caption{The poles trajectories of the $\hat a$ and $\hat c$ modes with the change of the compensation frequency, when we introduced the idler fields\,(after removing the instability at high frequency induced by pondermotive effect in the arm cavity). The arrows show the direction of compensation frequency increasing\,(under the condition of $\omega_s = G$). Using our sampling parameters, these two poles are both stable only when $\omega_{\rm comp}\in\omega_{\rm opt}+[91.39\rm Hz,91.76 \rm Hz]$. The system will be unstable around 2900\,{\rm Hz} or 3100\,{\rm Hz} if the compensation is out of this region\,(corresponding to the red points and blue points).} \label{fig:RLP_compensation_1}
\end{figure}

We plot the poles trajectories of the modes $\hat a$ and $\hat c$ on the complex frequency domain in Fig.\,\ref{fig:RLP_compensation_1}, when we tune the compensation frequency under the PT-symmetry condition $\omega_s=G$. We found that these two modes can both be stable only when the compensation frequency is within a small frequency domain around $\omega_{\rm opt}$.

\begin{figure}[h]
	\centering
		\includegraphics[width=0.9\linewidth]{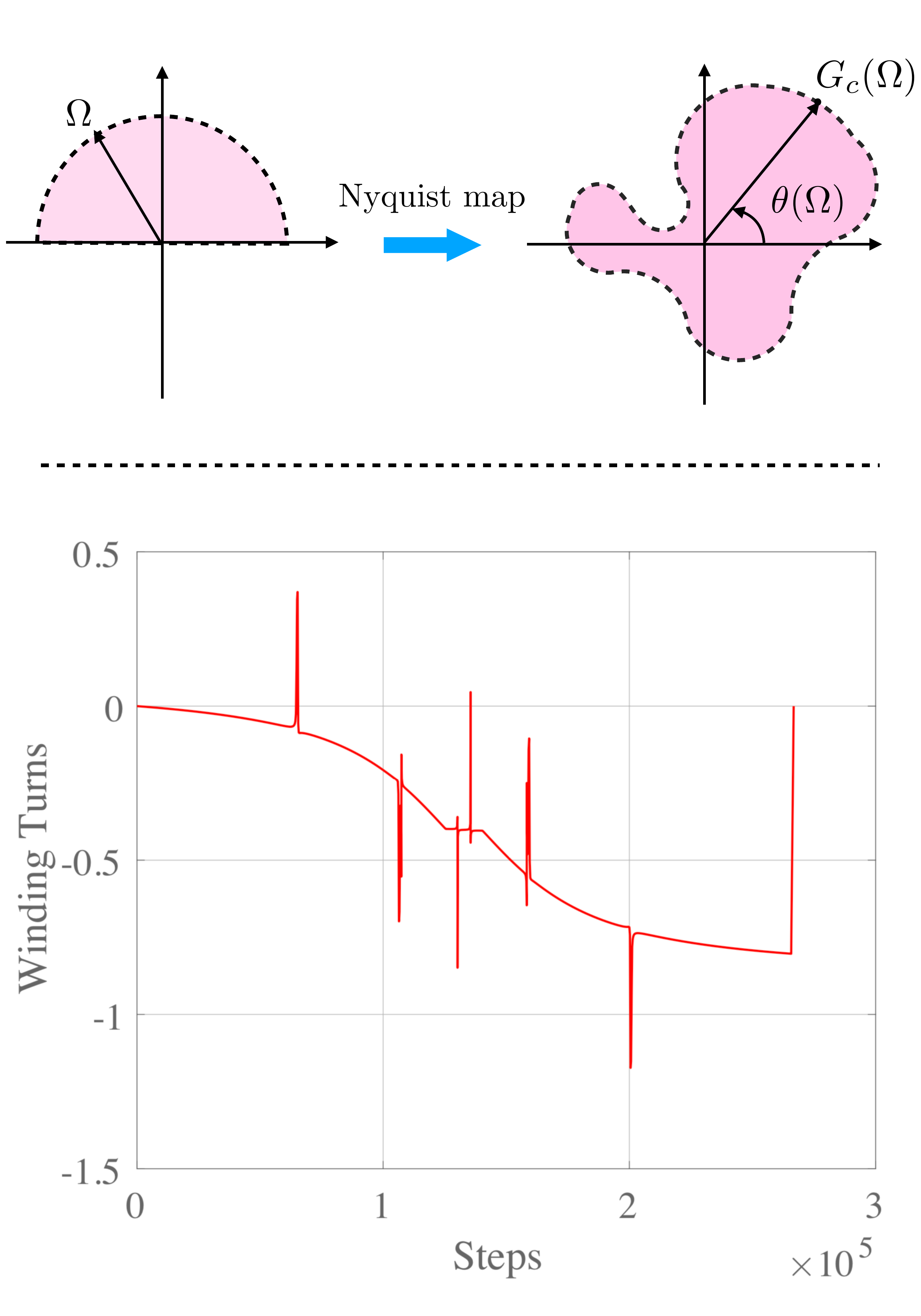}
	\caption{The winding number: the system is stable with no poles on the upper side of the complex plane\,(i.e. the winding number is zero) when we add a damping rate of 0.02\,Hz to compensate the optical spring instability. The contour in the complex $\Omega$-plane is chosen to be the upper half-circle with a radius equal to $2\pi\times2\times 10^4$\,rad/s. We homogeneously choose the data points in this half-circle contour and map it to the $G_c(\Omega)$-plane. The winding turns are defined as $\theta(\Omega)/2\pi$ explained in the upper pannel, where the shape peaks/dips indicates a close-pass of the contour near the origin.}
	\label{fig:winding-analysis} 
\end{figure}

The above discussions demonstrate that the instabilities which happen around the detuning frequency $\Omega=\Delta$ can be compensated by the careful tuning of the pumping frequency in the optomechanical filter cavity, while the optical spring instability at low frequency can be controlled by interferometer feedback servo system. For an overall verification, we also analyzed the stability using the Nyquist criteria\,\cite{1986AmJPh..54.1052P}, which is a conventional diagrammatic method for studying stability using the open-loop transfer function. In our analysis, the open-loop transfer function $G_{o}(\Omega)$ is chosen to describe the response of both the main detuned signal-recycling interferometer and the pondermotive interaction in the optomechanical filter cavity, shown in Fig.\,\ref{fig:schematic_setup},  while the SRM treated as the feedback kernel. The close loop transfer function is therefore given by 
\be
G_c(\Omega) = \frac{G_o(\Omega)}{1-r_{\rm SRM}G_o(\Omega)},
\ee
where those transfer functions are calculated using full-transfer matrix approach without making the resolved-sideband and single-mode approximations (see Section\,\ref{sec:5}).
The system is stable only if the contours generated by the Nyquist map from the complex $\Omega$-plane to the $G_c(\Omega)$-plane do not encircle the origin. In Fig.\,\ref{fig:winding-analysis}, we plot the winding numbers of the Nyquist contour around the origin of the complex $G_c(\Omega)$-plane when the above frequency tuning is performed. It finally tells us that the system is stable since the winding number to the zero point is zero in the case when we add an extra damping rate equal to $0.02$\,Hz to damp the test mass optical spring instability\,(using the sampling parameters given in Tab.\,\ref{tab:parameters}), under the condition that the filter cavity pumping frequency is carefully tuned.

\section{Full analysis of an optomechanical realisation using exact transfer matrix approach}\label{sec:5}
In the discussion above, we have used the following approximations: (1) single-mode approximation for each optical system so that  $\Omega L_{\rm arm}/c\ll1$ and $2\omega_mL_{\rm SRC}/c\ll1$, (2) the resolved sideband approximation that allows to ignore the idler mode in the parametric process. In this section, we show the exact case, especially to make use of the ignored idler mode to compensate for the pondermotive degradation of sensitivity. Our numerical calculation of the sensitivity curve follows the standard transfer matrix approach, which was briefly summarised in\,\cite{Li2021enhancing}.  

The method of Wiener filtering can be used to combine the signal and idler channels. The detailed derivation of this multi-channel Wiener filtering method is discussed in the appendix of\,\cite{Rehbein2008}. Denoting the two Wiener filter functions as $K_{1,2}(\Omega)$, and the combined output $\hat y(\Omega)$ can be written as:
\be
\hat y(\Omega)=K_1(\Omega)\hat y_s(\Omega)+K_2(\Omega)\hat y_i(\Omega),
\ee
where $\hat y_{s/i}$ is the measured signal/idler field quadrature operator. These operators can be formally represented as:
\be
\hat y_s=\mathbf{n}_s\cdot \mathbf{a}_s+d_sh,\quad \hat y_i=\mathbf{n}_i\cdot \mathbf{a}_i+d_ih,
\ee
where $\mathbf{n}_{s/i}$ and $d_{s/i}$ are the transfer functions for noise field $\mathbf{a}_{s/i}=(a_{1s/i},a_{2s/i})^T$ and the GW signal $h$.

\begin{figure}[th]
\centering
\includegraphics[width=0.48\textwidth]{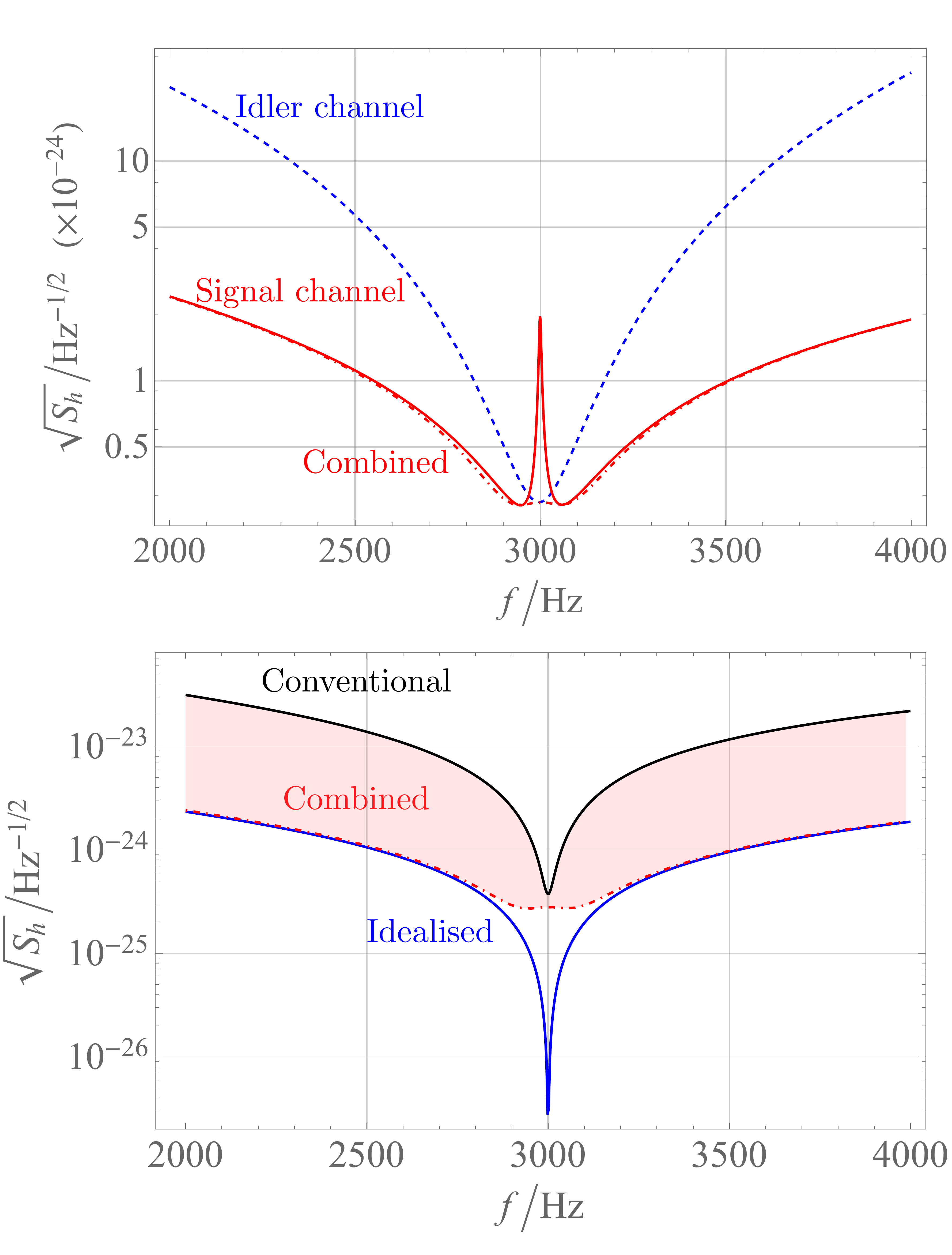}
\caption{The quantum-noise-limited sensitivity around 3000\,Hz obtained by the full transfer matrix simulation. Upper panel: The detector sensitivity obtained from the signal/idler channels and their Wiener-filtering combination. Lower panel: comparing the Wiener-filtering combined sensitivity curve with that of the conventional detector and idealized PT-symmetric configuration.}\label{fig:wiener}
\end{figure}

\begin{figure}[th]
\centering
\includegraphics[width=0.48\textwidth]{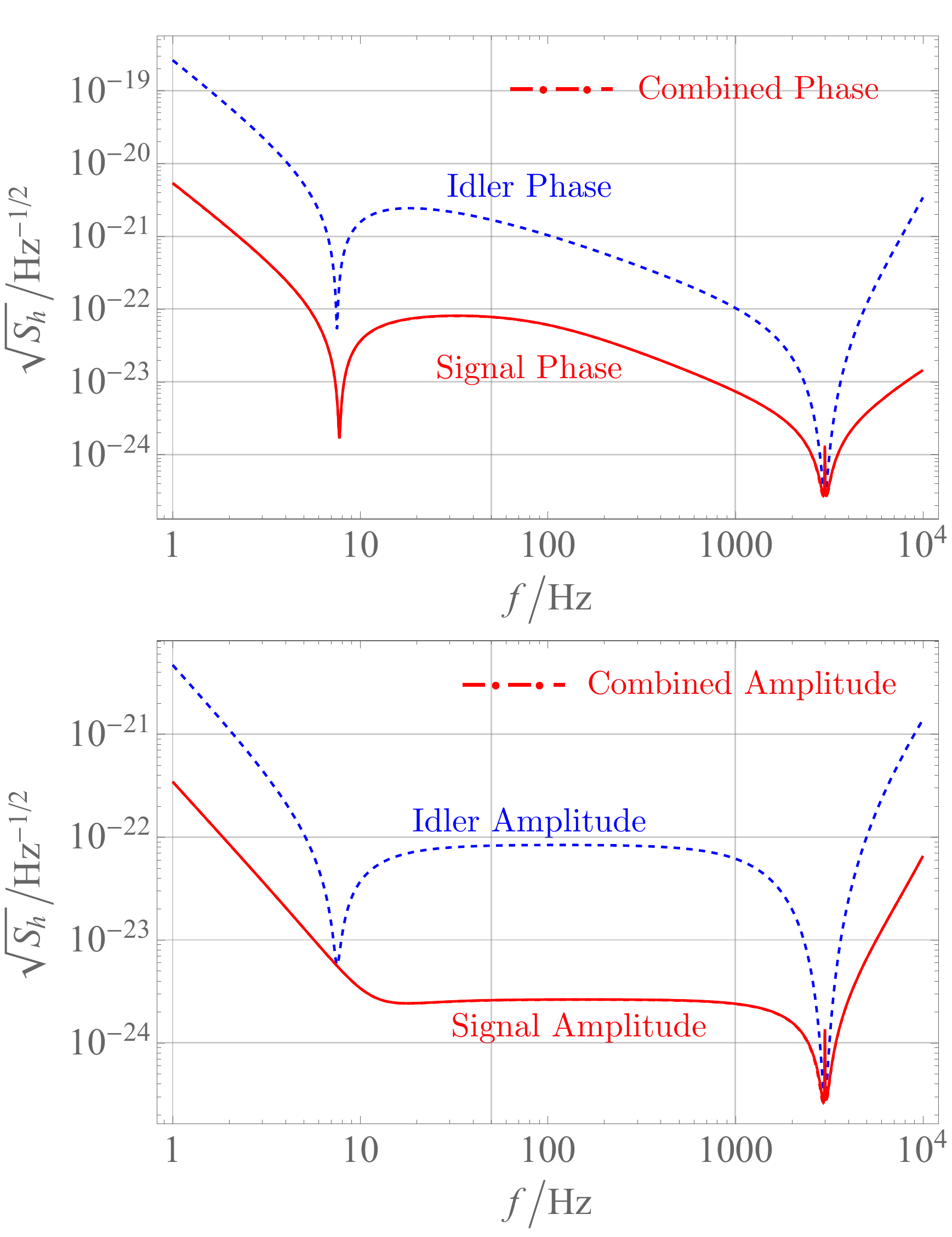}
\caption{The quantum-noise-limited sensitivity obtained by the full transfer matrix simulation in frequency band [$1$,$10^4$]\,Hz. Upper panel: The detector sensitivity obtained from the signal/idler channels (phase quadrature) and their Wiener-filtering-combination. Lower panel: The detector sensitivity obtained from the signal/idler channels (amplitude quadrature) and their Wiener-filtering-combination. Note that the dashed line is almost overlapped with the signal channel except at around 3000\,Hz, which can be hardly distinguished in the broad frequency band.}\label{fig:global_wiener}
\end{figure}

Optimizing the sensitivity by varying these two filter functions leads to the minimum noise given as the inverse of the largest eigenvalue of the following matrix:
\be
\mathbf{M}=\mathbf{N}^{-1}\cdot\textbf{$\Sigma$}.
\ee
where the noise and signal matrix are given by:
\be
\mathbf{N}=
\left [
\begin{array}{cc}
\mathbf{n}_s\mathbf{S}_{a_sa_s}\mathbf{n}_s^\dag&\mathbf{n}_s\mathbf{S}_{a_sa_i}\mathbf{n}_i^\dag\\
\mathbf{n}_i\mathbf{S}_{a_ia_s}\mathbf{n}_s^\dag&\mathbf{n}_i\mathbf{S}_{a_ia_i}\mathbf{n}_i^\dag
\end{array}
\right],
\ee
and
\be
\Sigma=
\left [
\begin{array}{cc}
|d_s|^2&d_sd_i^*\\
d_id_s^*&|d_i|^2
\end{array}
\right].
\ee
The combined sensitivity is shown in Fig.\,\ref{fig:wiener}, where the Wiener filtering process favours the idler channel near 3\,kHz. This combined sensitivity has a much larger bandwidth and sensitivity than the conventional configuration at around 3\,kHz. Moreover, we also provide the behaviour of the sensitivity curve in a larger frequency band, say $[1,10^4]$\,Hz, in Fig.\,\ref{fig:global_wiener}. Besides, the Wiener filtering combined sensitivity has almost no difference from the signal channel since the signal channel is favoured outside the frequency band around 3\,kHz. The low-frequency peak corresponds to the optical spring effect due to the main interferometer detuning. 

\begin{table}[h]
        \centering
       \begin{tabular}{l c c}
         \hline\hline
        Parameters & Symbols & Values  \\
        \hline 
        Arm cavity length&$L_a$& 4\,km \\
           \hline
        ITM mirror transmissivity&$T_{\rm ITM}$& 0.028 \\
           \hline
        Arm cavity power&$P_c$& 1.5\,MW \\
          \hline
        Arm cavity loss&$\epsilon_{\rm ETM}$& $10$\,ppm \\
          \hline
        Arm cavity mirror mass&$M$& $100$\,kg \\
          \hline
        iSRM transmissivity &$T_{\rm iSRM}$& $573.36$\,ppm \\
        \hline
        SR cavity length &$L_c$& 56\,m\\
           \hline
        SR cavity loss& $\epsilon_{\rm SRC}$& $150$\,ppm\\
        \hline
        Oscillator mass &{$m$} & {$10$\,mg} \\
          \hline
        Mechanical frequency &{$\omega_m$} & {$0.1$\,MHz} \\
           \hline
        Mechanical quality factor &{$Q_m$} & $10^{10}$\,\cite{Galliou2013} \\
          \hline
        SRM transmissivity&$T_{\rm SRM}$  &  $0.02$  \\
         \hline
        Filter cavity round trip loss&$\epsilon_{ f}$  & $10$\,ppm  \\
         \hline
        Filter cavity power&{$P_f$} & 7481.25\,W\\
          \hline
         Filter cavity length&{$L_f$} & $40$m\\
          \hline
          Filter cavity environmental temperature &$T_{\rm en}$  &  $1$\,K  \\
         \hline
          Input squeezing level &$r_{\rm sq}$ &  $10$\,dB  \\
         \hline\hline
       \end{tabular}
       \caption{Sample parameters for  the detuned PT-symmetric gravitational wave detector.}\label{tab:parameters}
\end{table}

\begin{figure}[h!]
\centering
\includegraphics[width=0.48\textwidth]{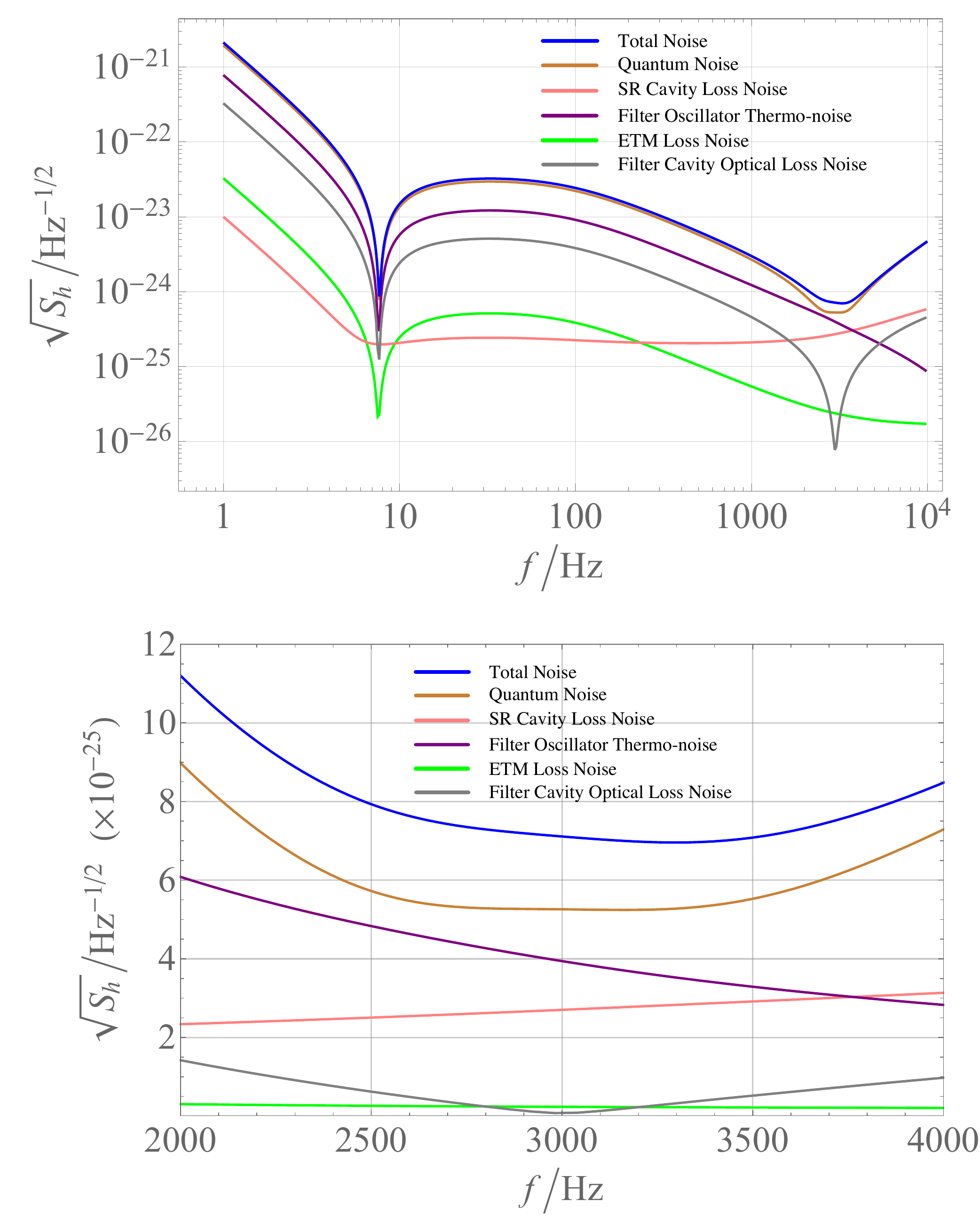}
\caption{The noise budget of the detuned PT-symmetric gravitational wave detector, where the noise due to the signal recycling cavity loss, filter cavity loss, ETM loss and the thermal noise of the mechanical oscillator in the filter cavity is considered. The upper panel is the noise budget from 1\,Hz to $10^4$\,Hz, while the lower panel is the noise budget around 3000\,Hz.}\label{fig:noise_budget}
\end{figure}

Furthermore, we also considered the effect of various losses in this protocol, such as the noise introduced by the optical loss in the signal recycling cavity (formed by iSRM, beam-splitter and the two ITMs in Fig.\,\ref{fig:schematic_setup}), the optical loss in the arm-cavity (which is attributed to the ETM loss) and the optical-loss/thermal noise in the optomechanical filter cavity. The corresponding sampling parameters are listed in Tab.\,\ref{tab:parameters}. The resultant noise budget is shown in Fig.\,\ref{fig:noise_budget}, where the sensitivity is plotted in both the frequency range of $1-10^4$\,Hz and around 3000\,Hz. As also discussed in\,\cite{Miao_kHz_2018}, the thermal noise in the filter cavity behaves similarly to the optical loss in the filter cavity in the broad frequency range, while a bit differently around the detuning frequency $\Omega=\Delta$ due to the suppression of the filter's optical loss noise near the detuning frequency\,\cite{miao2015enhancing}. The noise components of the noise budget in Fig.\,\ref{fig:noise_budget} are plotted using the Wiener combination coefficient computed based on the total noise of idler and signal channels, and we also assumed an injection of 10\,dB phase squeezing field.

\section{Astrophysical implications}\label{sec:6}
\begin{figure}[h]
    \centering
    \includegraphics[width=0.48\textwidth]{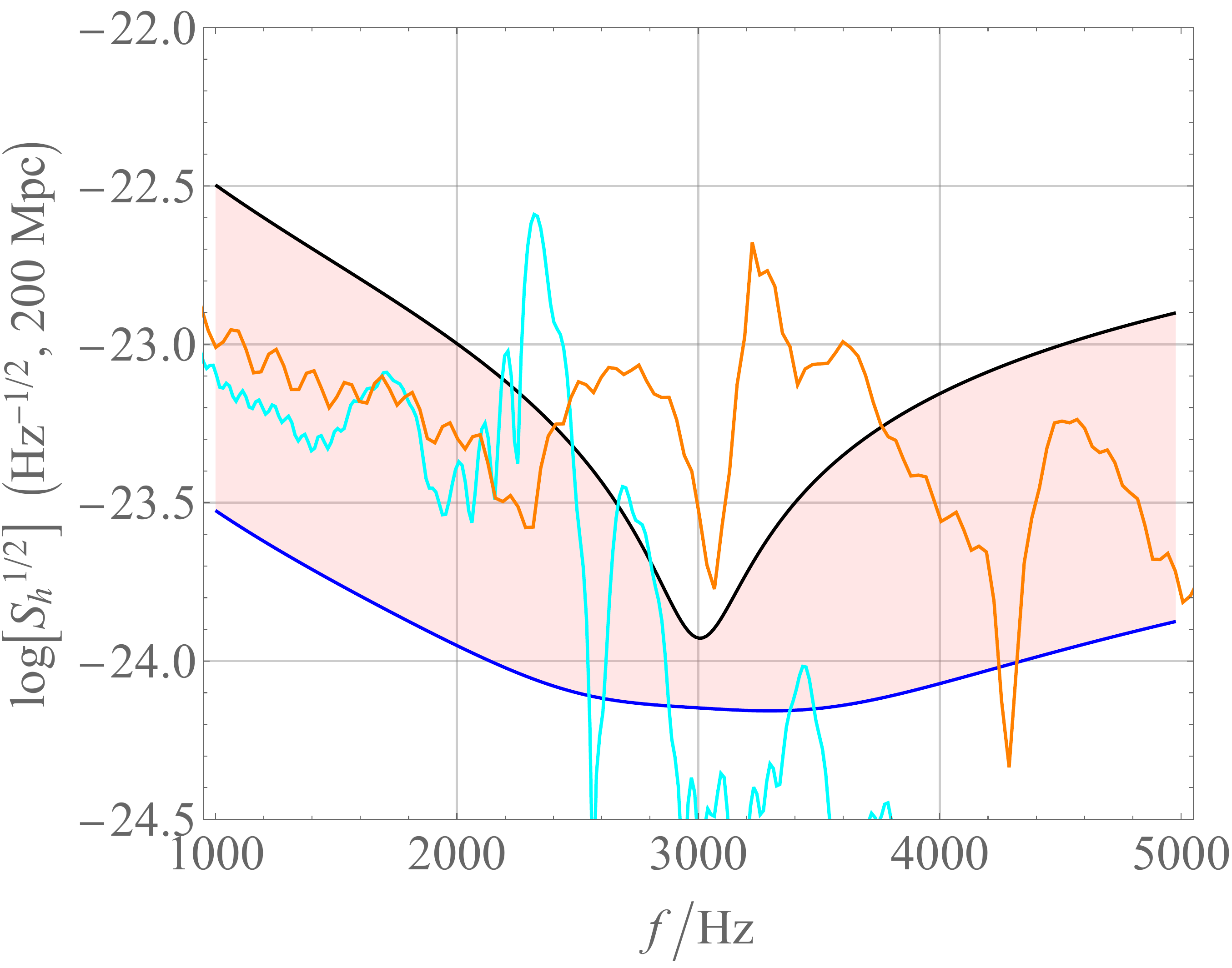}
    \caption{The sensitivity curves with GW signal emitted by neutron star merger remnants with two different equation of states. The black line is the sensitivity of the conventional detuned interferometer, while the blue line is the sensitivity of our protocol. These sensitivity curves are plotted using the parameters in Tab.\,\ref{tab:parameters}. The orange line is the BNS merger waveform of the APR4-q10-M1375 model calculated in\,\cite{PhysRevD.91.064001} and the skyblue line is the waveform of 15H model listed in the data bank\,\cite{Waveform-web}.
    }\label{fig:signal-noise}
\end{figure}

GW observations of binary neutron star (BNS) inspiral have already been achieved and various constraints on the equation of state (EoS) of neutron stars (NS) have been interpreted from the observation of GW170817 \citep{Abbott2018tidal,Shibata2017,Bauswein2017,Ruiz2017,Shibata2019,Rezzolla2017}. Boosting the sensitivity of GW detectors around the frequency of 3\,kHz could open up the possibility of direct detection of the merger and post-merger GW signals of BNS merger events, with which the understanding of the EoS of dense matter at supranuclear densities could be pushed to a new level. 

Based on BNS merger simulations carried out by various groups, it is widely accepted that the fate of the post-merger remnant is determined by the ratio between the maximum mass of a cold non-rotating NS (i.e., $M_\mathrm{TOV}$) and the total mass of the binary (which could be accurately measured by inspiral GW signal) \cite{Margalit2019}. Therefore, simply a non-/detection of the GW signal from the post-merger remnant could tell that the remnant experiences prompt/delayed-collapse to a black hole (BH). This already allows for an independent constraint on $M_\mathrm{TOV}$. Furthermore, it has been found that the post-merger GW signals could be used to constrain crucial properties of the merging NSs such as compactness and tidal deformability \cite{Bauswein2012b,Hotokezaka2013,Takami2014,Bernuzzi2015a,Bauswein2015b}. However, the most relevant frequency range of the post-merger GW signals lies in the range from 2\,kHz to 3\,kHz, which is too high for the current generation GW detectors to resolve even for close sources such as GW170817. Our design could make it possible for a measurement of the frequency peaks in the post-merger phase and hence make complementary constraints on NS properties.

Moreover, the density of the BNS merger remnant could be several times higher than the inspiral stage. Strong interaction phase transition is suggested to be possible under such conditions and could leave detectable features in the post-merger GW signals \cite{Bauswein2019,Most2019}. Compared the cases without a phase-transition, the peak frequency of the post-merger GW is found to shift to a even higher value. Having a detector with a sufficiently broad sensitivity and frequency resolution at roughly 3\,kHz could identify such a shift and constrains the density range of the strong interaction phase transition, see Fig.\,\ref{fig:signal-noise}. 

\section{Dicussion and Conclusion}
In this work, we provided an alternative design protocol based on the PT-symmetry to the high-frequency gravitational wave detector, targeted at the physics of binary neutron star coalescence. We
have analyzed in detail the optomechanical realization of a detuned PT-symmetric interferometer, targeted at improving the sensitivity to the kilo-Hertz GWs using single-mode approximation and full transfer matrix simulation. The effects of the idler field and pondermotive interactions are analyzed in the single-mode approximation, which provides physical insight for understanding the sensitivity curves obtained using transfer matrix simulation. We showed that using the same parameter setting as in the tuned case, after (1) compensating the optical spring in the optomechanical quantum amplifier and (2)  the Wiener-filtering combination of the measurement data obtained from both signal channel and idler channel, we can in principle achieve a significant boost of the sensitivity around $3$\,kHz. This sensitivity is 10-times better than the conventional design. The dynamical instability of the system is induced by the optical spring effect, which can be controlled in the detector feedback servo system.  Our work focuses on the conceptual designs of this protocol while leaving the more practical and technical considerations to future studies.

\section{Acknowledgements}
The authors thank our colleagues AEI-Caltech-MSU MQM discussion group and also Huan Yang for fruitful discussions. Y. M. is supported by the start-up fund provided by Huazhong University of Science and Technology. C. Z. acknowledges the support of Australian Research Council through Grants No. DP170104424 and No. CE170100004. X. L. and Y. C.’s research is funded by the Simons Foundation (Grant No. 568762). H. M. acknowledge support from the Birmingham Institute for Gravitational Wave Astronomy and the UK EPSRC New Horizons award (Grant No. EP/V048872/1). H. M. has also been supported by a UK STFC Ernest Rutherford Fellowship (Grant No. ST/M005844/11), and the start-up fund provided by Tsinghua University and the State Key Laboratory of Low Dimensional Quantum Physics of Tsinghua University.

\bibliography{reference}
\end{document}